\documentclass[titlepage,11pt,twoside]{article}

\usepackage[normalem]{ulem}
\usepackage{graphicx}
\usepackage{tikz}
\usetikzlibrary{calc,shapes,backgrounds, arrows, positioning}
\usepackage[myheadings]{fullpage}
\usepackage{pmetrika}
\usepackage{pmbib}
\usepackage{IEEEtrantools}
\usepackage{amsmath}
\usepackage{amssymb}
\usepackage[natbibapa]{apacite}
\usepackage{float}
\usepackage{todonotes}
\usepackage{submit}
\usepackage{enumitem}
\usepackage{multirow}
\usepackage{algorithm}
\usepackage{algpseudocode}
\usepackage{url}
\usepackage[pagewise]{lineno}
\setlist{nosep}

\tikzstyle{item}=[rectangle,draw=black,fill=white,
                    font=\scriptsize\sffamily\bfseries,inner sep=6pt]
\tikzstyle{factor}=[ellipse,draw=black,fill=white,
                     font=\scriptsize\sffamily\bfseries,inner sep=-1pt]

\tikzstyle{arr}=[-latex,black,line width=1pt]
\tikzstyle{doublearr}=[latex-latex,black,line width=1pt]

\begin{document}
\begin{titlepage}

\linespacing{1}

\title{Ordinal Outcome State-Space Models for Intensive Longitudinal Data}

\author{Teague R. Henry \\ Department of Psychology and School of Data Science, University of Virginia \\
  Lindley R. Slipetz \\ Department of Psychology, University of Virginia \\
  Ami Falk \\ Department of Psychology, University of Virginia\\
  Jiaxing Qiu \\ School of Data Science, University of Virginia\\
  Meng Chen \\ Health Sciences Center, University of Oklahoma}

\markboth{Psychometrika}{*}

\vspace{\fill}
\vspace{\fill}
\vspace{\fill}

\linespacing{1}\fontsize{8}{10}\selectfont

\end{titlepage}\vspace*{24pt}

\linespacing{1}

\begin{center}\vskip3pt

\vspace{32pt}

Abstract\vskip3pt

\end{center}

\begin{abstract}
Intensive longitudinal (IL) data are increasingly prevalent in psychological science, coinciding with technological advancements that make it simple to deploy study designs such as daily diary and ecological momentary assessments. IL data are characterized by a rapid rate of data collection (1+ collections per day), over a period of time, allowing for the capture of the dynamics that underlie psychological and behavioral processes. One powerful framework for analyzing IL data is state-space modeling, where observed variables are considered measurements for underlying states (i.e., latent variables) that change together over time. However, state-space modeling has typically relied on continuous measurements, whereas psychological data often comes in the form of ordinal measurements such as Likert scale items. In this manuscript, we develop a general estimating approach for state-space models with ordinal measurements, specifically focusing on a graded response model for Likert scale items. We evaluate the performance of our model and estimator against that of the commonly used ``linear approximation'' model, which treats ordinal measurements as though they are continuous. We find that our model resulted in unbiased estimates of the state dynamics, while the linear approximation resulted in strongly biased estimates of the state dynamics. 
\end{abstract}

\vspace{\fill}

\newpage

\section{Introduction}

Intensive longitudinal data in psychological and behavioral sciences typically consists of self-report, behavioral and/or psychophysiological data collected multiple times a day for multiple participants. There are a number of study designs that produce intensive longitudinal data, such as daily diary studies, experience sampling, and ecological momentary assessment. In this manuscript, we develop the estimation approach for modeling intensive longitudinal data in the state-space framework while accounting for the presence of ordinal measurements such as Likert scale-type items. Our motivating example of a study design that produces intensive longitudinal data is that of ecological momentary assessment, a design that is increasingly used in the study of psychopathology, however, we note that the state-space approach is appropriate for the modeling of any intensive longitudinal data.

Ecological momentary assessment (EMA) is a type of intensive time series in which data are gathered in the participants’ natural environments, typically multiple times a day over several days. This results in strong ecological validity of measurement as, in theory, the data reflects the participants' lived experience rather than the effects of a lab. In addition, a benefit of EMA is that it has the ability to capture the near real-time dynamics of the phenomenon under study, rather than offering mere snapshots, as is the case with laboratory-based designs or more traditional longitudinal designs \citep{wright_relationship_2021}. The dynamic relationships of psychological variables and behavior can be tracked in near real time in EMA. For instance, consider the relationship between stress and cigarette smoking. In a laboratory study, the participant would be required to recall when they were stressed, when they smoked, and if there was a link between the two \citep{mote_ecological_2020}. For EMA, the need for retrospective recall is eliminated as stress and cigarette smoking can be tracked in near real time via smartphone-based EMA data collection. EMA studies of psychopathology have been increasing in the last few years \citep[][]{stumpp2023, saulnier_cognitive_2022, seidman_ecological_2022} due to both the increased ubiquity of smartphones and the theoretical turn towards complex systems depictions of psychopathology. 
	
EMA data can be modeled in a variety of ways from observed variable time series models to mixed effects models. Here, we use a general framework that represents observed time-varying variables as indicators of underlying, latent state variables: \textit{the state-space model} \citep{durbin_time_2012}. Using this framework/model, the temporal dynamics of the psychological phenomenon under study are wholly determined by the temporal relations between the states, which are then ``measured'' by the observed variables. As a general framework, state-space models can allow for non-linear relations and unequal time intervals between data collections, but the key benefit of the state-space approach relevant to the work presented here is the ability to account for different measurement types. State-space modeling has been used to study psychological phenomena previously, and we refer readers to the works of \citet{Oud1999,browne2005representing,Chow09c,Song2009} as representative methodological work in this space.

EMA data can be collected in a variety of modalities from psychological batteries \citep{levinson_personalizing_2023} to physiological/biological assay data \citep{ditzen_positive_2008}.  One common type of data collected in EMA studies (and more broadly in psychological or behavioral studies) are ordinal responses. Generally speaking, an ordinal variable is one where the values of the variable can be rank-ordered, but the numerical distance between the values is not defined. The best way of illustrating this is via example. One common example of an ordinal variable is a Likert scale item, which, with 5 responses might have the following options: 1 - Strongly Disagree, 2 - Disagree, 3 - No opinion, 4 - Agree, 5 - Strongly Agree. While the numerical values can be ordered based on the strength and direction of the agreement, the theoretical distance between 1 - Strongly Disagree and 2 - Disagree is not necessarily the same distance between 2 - Disagree and 3 - No opinion. Similarly, another ordinal type item would be a binned count of substance use: 0 - No use, 1 - 1-2 drinks, 2 - 3-5 drinks, 3 - 5+ drinks. Here, the distances between the response categories are prima facie unequal and ill-defined. The issue with ordinal items is, of course, that they are not continuous. Most analytic methods commonly used in psychological science treat ordinal variables as continuous (indeed, the above references for state-space modeling applied to psychological data all make use of this approximation), and previous work on the quality of this approximation has shown that this results in decreased true positives, increased false positives, and biased effect size estimates in a simulation study where parameter values were known \citep{liddell_analyzing_2017}. In this manuscript, we develop a general estimating approach for state-space models with non-continuous outcomes, and show how this estimating approach can be used to fit state-space models with ordinal outcomes using graded response model measurements. We evaluate the performance of our approach in a simulation study against that of the linear approximation approach that is commonly used in analyzing psychological data.

The remainder of this paper is structured as follows: we begin with an overview of the state-space modeling framework and discuss specific technical issues that arise during model estimation. We then describe the ordinal outcome state-space model that is the core contribution of this manuscript. Following that, we describe the four central issues and solutions that make estimating these models feasible: model identification, state filtering, parameter estimation and standard error approximation. 

Following the technical description of the model and its estimation methods, we present a simulation study that evaluates the performance of the proposed estimating method, as well as evaluates how well a linear approximation approach (i.e. what is most typically used when analyzing ordinal psychological data) can recover state dynamics. We close with a discussion of the modeling approach and its current limitations, as well as a discussion of future directions in this methodological space. 

\section{State-Space Models}

State-space modeling is a general framework for representing a dynamical process of latent variables (states) which have observed variables as their measurements \citep{durbin_time_2012,kalman_new_1960}. While the state-space modeling framework can be seen as related to dynamic structural equation modeling, particularly when both the latent states and the measured variables are normally distributed with linear dynamics \citet{asparouhov_dynamic_2018}, the state-space approach is a broader framework than dynamic structural equation modeling. State-space models allow for, in theory, non-linear dynamics and non-normally distributed states/ measurements \citet{durbin_time_2012}, though most current implementations of state space models lack the same capability as DSEM for simultaneous multi-participant models.

A simple example of a state-space model is that of a discrete-time, time-invariant, normal-state-normal-measurement with linear state dynamics, which can be described with the following equations:
\begin{align}
    \mathbf{x}_{t+1} &= \mathbf{A}\mathbf{x}_t + \mathbf{\varepsilon}_t \\
    \mathbf{y}_{t} &= \mathbf{C}\mathbf{x}_t + \mathbf{\varsigma_t}\\
    \mathbf{\varepsilon}_t &\sim N_p(0, \boldsymbol{\Sigma}) \\
    \mathbf{\varsigma}_t &\sim N_q(0, \boldsymbol{\Psi}) 
\end{align}
where $\mathbf{x}_t$ is a $p$ dimensional vector of state values at time $t$, $\mathbf{y}_t$ is a $q$ dimensional vector of observed measurements at time $t$, $\mathbf{A}$ is a $p \times p$ matrix that governs the dynamics of the states over time, $\mathbf{C}$ is a $q \times p$ matrix of state-measurement loadings, $\boldsymbol{\varepsilon}$ is the multivariate normally distributed \textit{disturbance} term with mean 0 and covariance matrix $\boldsymbol{\Sigma}$, and $\boldsymbol{\varsigma}$ is the multivariate normally distributed \textit{measurement error} term with mean 0 and covariance matrix $\boldsymbol{\Psi}$. 

To unpack the description of this model as ``discrete time, time invariant, normal state-normal measurement with linear state dynamics'': Discrete time refers to the equally spaced intervals with respect to $t$. In a discrete time state-space model, the time interval between measurements is assumed to be constant (and therefore ignorable). Time invariant refers to the various parameter matrices ($\mathbf{A}, \mathbf{C}, \boldsymbol{\Sigma}, \boldsymbol{\Theta}$) not varying as a function of $t$. Normal state-normal measurement corresponds to the use of multivariate normal disturbance and measurement error distributions, while the linear state dynamics refer to the use of $\mathbf{A}\mathbf{x}_t$ as the state dynamics.

It is important to note that each one of these assumptions can, in theory if not in practice, be relaxed. State-space models can be proposed in continuous time as differential equations, allowing for unequally spaced intervals. All parameters can be made to be time-varying, while states and measurements can have non-normal marginal distributions. Finally, non-linear state dynamics are also possible to represent in the state space modeling framework (e.g., instead of linear multiplication $\mathbf{A}x_t$, the state dynamics can be a nonlinear function $f(\mathbf{A},x_t)$).

However, as with any complex modeling approach, the devil is in the (estimation) details. State-space modeling originated in an engineering context \cite{kalman_new_1960}, with some of the first public use cases being for the navigation and control of the Apollo space mission \citet{mcgee_discovery_1985}. In an engineering context, the model parameters are often considered fixed and known, and the interest is in online estimating the trajectories of the states in real time. Additionally, within the engineering context the measured variables, while not necessarily normally distributed, tend to be continuously distributed, which simplifies estimation considerably. The estimation of the state values is known as prediction, filtering or smoothing, depending on if the states are estimated using only previous state estimates, previous state estimates and current measurements, or all state estimates and measurements past and future, respectively. In the context of parameter estimation, filtering is the most relevant, and we will focus our discussion on filtering in this manuscript.

The first filtering approach, the Kalman filter \cite{kalman_new_1960}, was developed for models of the form described in Eqs. 1-4, and provides an analytic solution to estimating the expected value of states at each timepoint. Since its development, the Kalman filter has been extended to continuous time models \citep[Kalman-Bucy;][]{kalman_new_1961}, non-linear state dynamics \citep[Extended Kalman][]{sorenson_kalman_1985}, non-linear measurement functions \citep[Unscented Kalman][]{wan_unscented_2000}, and regime-switching state space models \citep[Kim-Nelson Filter][]{kim_state-space_2017}, to name just a few of the variants. It is important to note here that all of these analytic filters assume that the measurements are continuous. Filtering when measurements are discrete/categorical requires the use of a different class of methods, which will be discussed later in this manuscript.

Parameter estimation (or system identification as it can be known) is a rarer need in most common applications of state space models, but is extremely important when applying state space models to psychological or behavioral data. In the case of the model described in Eqs. 1-4, the Kalman filter permits a prediction error decomposition approach to calculating the likelihood directly. This in turn allows for gradient-based optimization, however relaxing the various assumptions of that model quickly results in intractable likelihood expressions \citep{durbin_time_2012}. Additionally, state-space models tend to have ill-behaved likelihood surfaces due to weakly identified parameters \citep{ionides_inference_2015}, making gradient-based approaches to optimization difficult, if not impossible, outside of a restricted subset of model types. 

\section{Ordinal Measurement State-Space Models}

As stated previously, the issue with ordinal measurements in a state space modeling framework is that they are not continuous. This in turn makes most analytic filtering approaches, such as traditional Kalman filtering and the aforementioned variants, inappropriate. Applying standard approaches for fitting state-space models to ordinal outcomes is analogous to the use of fitting linear regression models to dichotomous data (and for that matter, fitting linear regression models to Likert type data). In the current manuscript, we refer to the use of continuous outcome state-space methods on ordinal data as the \textit{linear approximation} approach, and note that this linear approximation approach is used in many applications of state space modeling to psychological data.

Before developing our framework for ordinal measurement state space models, there are several previous approaches to estimating state-space models for ordinal data that need to be discussed. Broadly, previous work in this area falls into two categories: Bayesian-based methods and the work of van Rijn \citeyear{van_rijn_categorical_2008} which extends the work of \citet{fahrmeir_multivariate_2001}.

State-space models with ordinal/categorical outcomes have been proposed under a Bayesian estimating framework by a number of authors \citep{wang_bayesian_2013,chaubert_multivariate_2008}. The Bayesian approach permits flexible measurement distributions as well as arbitrary state dynamics (up to the limits of identification) using MCMC-based estimating techniques such as Metropolis-Hastings and/or conditional Gibbs sampling. However, the cost of this flexibility is in computation time and the need to specify priors for each parameter. This combination of computational needs with the need for expertise in Bayesian model specification makes the application of state-space modeling with ordinal outcomes difficult for the applied scientist. That being said, Bayesian estimation allows for bespoke models to be constructed and estimated in situations where frequentist estimation has difficulties, and the computational aspects of Bayesian estimation might be relieved with more efficient sampling methods such as Hamiltonian MCMC \citep{bickel_monte_1996} and/or variational methods \citep{blei_variational_2017}. As the goal of the current manuscript is to develop a standard frequentist estimation framework for these models, we will put the prior work on Bayesian estimation aside while noting its usefulness and applicability for this general class of models. 

Related to Bayesian estimation of state space models is the work of \citet{asparouhov_dynamic_2018} who have developed an approach to fitting dynamic structural equation models (DSEM). DSEM can account for ordinal measurements by  using probit measurement models. Probit measurement models assume that the underlying latent state is normally distributed, and is mapped onto the ordinal response categories via hard thresholds (e.g. if the state is below $.5$, $y = 0$, above $.5$ $y = 1$). The approach that Asparouhov and colleagues describes makes use of Bayesian estimation methods in Mplus \citep{muthen_mplus_1998} and additionally allows for the analysis of multiple participant's data via random effects. However, the probit model has hard thresholds, which in turn assumes that there is no measurement error in the response process (i.e. once the participant's latent state is in a certain region, they will always respond with the same category.) Accounting for measurement error in ordinal responses is better accomplished adopting methodology from item response theory, which both the current work, and the work of \citet{van_rijn_categorical_2008} do.

The work of van Rijn \citeyear{van_rijn_categorical_2008} comes closest to the goals of the current manuscript. In that work, van Rijn applied the general estimating techniques from \citet{fahrmeir_multivariate_2001}, which allows for dynamical systems with exponential family measurement distributions, and is estimated using an iterative Kalman filtering strategy. van Rijn, like the current paper, focused on graded response type outcomes, which can be put in exponential family form. While this approach does provide a consistent estimating framework, it has a number of disadvantages: First, it requires that certain parameters, including the state dynamics (here, $\mathbf{A}$) to be fixed or known. This is limiting in an idiographic modeling context, as the interest is typically in the individual differences in the latent state dynamics. Second, the requirement of exponential family measurement distributions, while broad enough to cover many different ordinal item measurement distributions, removes from consideration a number of useful measurement distributions, most notably zero-inflated distributions such as the zero-inflated negative binomial or zero-inflated Poisson. Finally, the implementation of this technique requires analytic derivations for the measurement models, and there are no open source implementations of the approach available (it must be noted that the code from \citet{van_rijn_categorical_2008} is available from the author on request). On the other hand, our implementation of the ordinal state space model uses a measurement family-agnostic estimating method, and does not require model parameters to be fixed and known (after model identification has been achieved), and we offer a open-source implementation of the modeling framework \citep{falk_netlabuvagenss_2023}. 

 Given the state of research on state-space models with ordinal measurements, our goal with this project was to develop a general frequentist approach to fitting state space models with arbitrary measurement models. In this manuscript, we specifically develop our state space models for graded response \citep[GR; ][]{samejima_graded_1997} measurements, a model that links an ordinal response to an underlying continuous state via cumulative logistic functions. GR models were originally developed (as most item response theory models are) in an educational assessment framework, and the idea behind a GR model is to capture how differing levels of underlying ability correspond to different proportions of the item being correct, with higher ability corresponding to more of the item being correct. However, we note here that the estimating methods and methods for producing standard errors that we describe here are applicable to any measurement model, (for example, zero-inflated distributions), after identification considerations have been made. 

The GR outcome state-space model can be expressed in the following form:
\begin{align}
    \mathbf{x}_{t+1} &= \mathbf{Ax}_t + \boldsymbol{\varepsilon}_t \\
    \boldsymbol{\varepsilon}_t &\sim N_p(0, \Sigma)\\
    p(y_{it} > j | \mathbf{x}_t) &= \frac{1}{1+\exp[-\alpha_i (\delta_i(\mathbf{x}_t) - \beta_{ij})]}
\end{align}
where $\mathbf{x}_t$, $\mathbf{A}$, and $\varepsilon_t$ are as previously described; $y_it$ is the $i$th ordinal measurement item, 
$p(y_{it} > j | \mathbf{x}_t)$ is the probability that the response of $\mathbf{y}_{it}$ being greater than the $j$th response category given the value of the states at $t$; $\alpha_i$ is the discrimination parameter for item $i$, $\beta_{ij}$ is the $j$th threshold parameter for item $i$, and $\delta_i(\mathbf{x}_t)$ is a $p \rightarrow 1$ selector function that assigns each item to a single state. The use of a selector function here is for notational clarity, and corresponds to the assumption that items do not load onto multiple states. However, there is no explicit reason (beyond complexity) for items to not be indicators of multiple states, we simply make that assumption here to reduce computational complexity.

\begin{figure}[H]
    \centering
    \includegraphics[width=.99\textwidth]{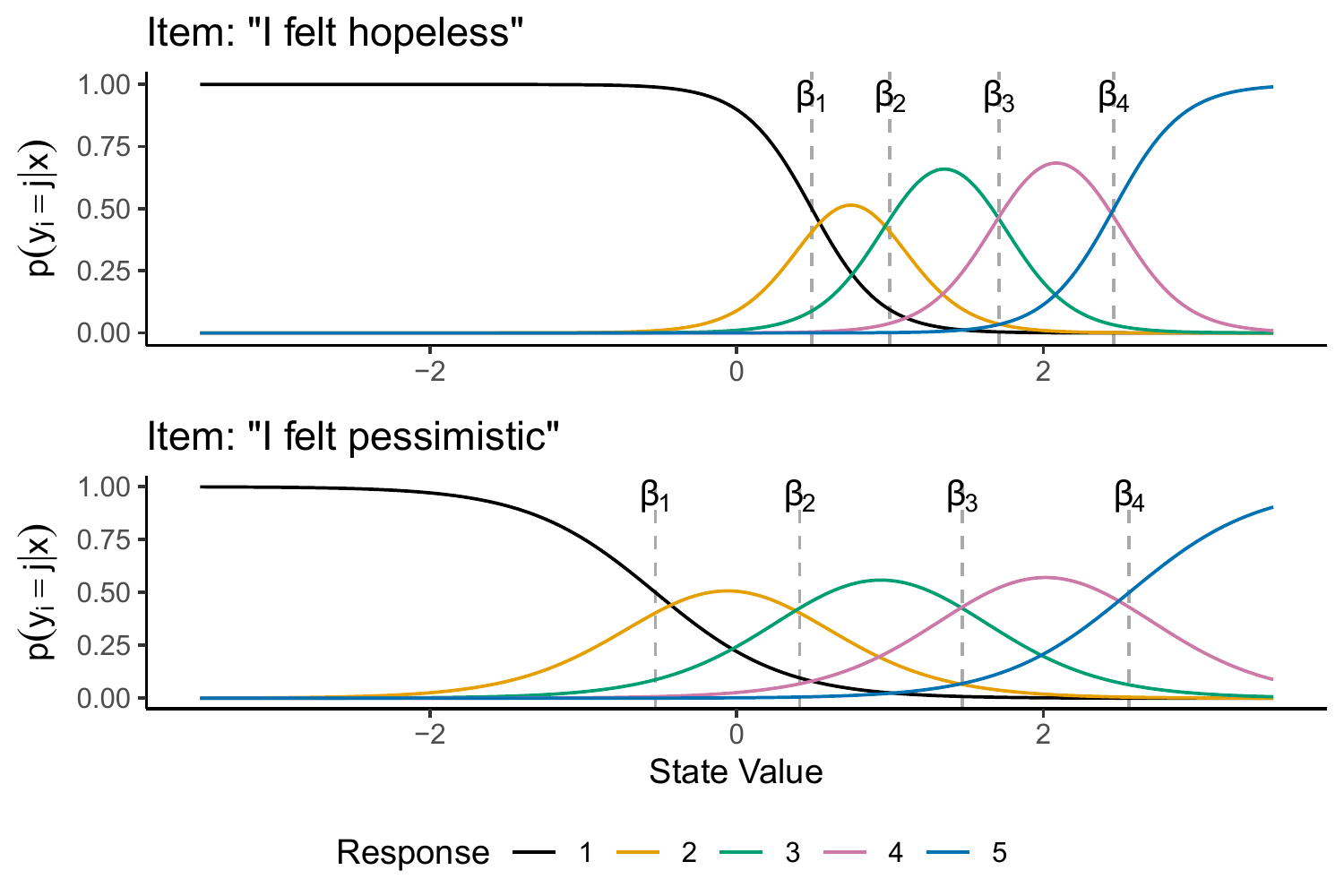}
    \caption{Item Response Probabilities for Two PROMIS Depression Items}
    \label{fig:GRM_examp}
\end{figure}

The probability of a response option $j$ for item $\mathbf{y}_{it}$ is then:
\begin{equation}
    p(y_{it} = j | \mathbf{x}_t) = p(y_{it} > j | \mathbf{x}_t) - p(y_{it} > j-1 | \mathbf{x}_t)
\end{equation}
with $p(\mathbf{y}_{it} > 0 |\mathbf{x}_t) = 1$.
Fig \ref{fig:GRM_examp} shows probability curves across values of $\mathbf{x}_t$ for two 5 response graded response items. These items were taken from the PROMIS Depression Battery \citep{pilkonis_item_2011}, which is an extensively validated and calibrated scale designed to collect high quality data from patients in a medical setting. The top panel shows the responses for the item ``In the past 7 days, I felt hopeless'', with item parameters $\alpha = 4.46$, and $\beta = [.49, 1.00, 1.71, 2.46]$, while the top panel shows the responses for the item ``In the past 7 days, I felt pessimistic'', with item parameters $\alpha = 2.38$ and $\beta = [-.53, .41, 1.47, 2.56]$.

One key aspect of graded response items is that different items are more or less better at measuring specific areas of the underlying scale. For example, the large $\alpha$ for the item "I felt hopeless", combined with the locations of that items $\beta$ values shows that "I felt hopeless" has the most optimal measurement when the underlying state is above around .5. The second item, "I felt pessimistic", has less discriminant measurement but captures a broader range of state values. 

\subsection{Identification and Estimation}

In developing the estimation strategy for the GR outcome state-space models described in Eqs. 5, 6, and 7, the first consideration we faced was how to identify the latent states. The \textit{identification problem} is a well-known issue in latent variable modeling from both a structural equation modeling and item response theory perspective, and refers to the fact that the location and scale of a latent variable is, by definition, arbitrary, whereas with observed data, the location and scale can be directly estimated \citep[i.e., scale indeterminacy;][]{baker_item_2004}. 

In cross-sectional models using latent variables, the identification problem is solved by fixing a factor loading to a specific value (usually 1) or by fixing a threshold in the probit approach to a specific value, which corresponds with the latent factor taking on the same scale as the indicator, or by fixing the variance of the latent variables to a constant (again, usually 1). Location is easily identified by setting the expected value of the factor to a constant (usually 0). 

Unfortunately, neither of these approaches to model identification can be easily applied to the identification of the GR state-space model, for the simple reason that the marginal distribution of the states is not directly parameterized by Eqs. 5 and 6. The marginal distribution of $\mathbf{x}_t$ for all values of $t$ is a function of the model dynamics $\mathbf{A}$ and the innovation covariance matrix $\boldsymbol{\Sigma}$. Specifically, $\mathbf{x}_t \sim N(0,\boldsymbol{\Gamma})$, where $\boldsymbol{\Gamma}$ can be calculated as follows:
\begin{equation}
    \text{vec}(\boldsymbol{\Gamma}) = (\mathbf{I}- \mathbf{A} \otimes \mathbf{A})^{-1} \text{vec}(\boldsymbol{\Sigma})
\end{equation}
where $\text{vec}(\cdot)$ is the vectorizing operator and $\otimes$ is the Kronecker product \footnote{This function is slightly different when dealing with continuous time models where the state dynamics are in differential equation form: $\dot{\mathbf{x}}_t = \mathbf{Ax}_t + \boldsymbol{\varepsilon}_t$. Here, the function is $\text{vec}(\boldsymbol{\Gamma}) = (\mathbf{A} \oplus \mathbf{A})^{-1} \text{vec}(\boldsymbol{\Sigma})$, where $\oplus$ is the Kronecker sum \citep{vatiwutipong_alternative_2019}.}. 

It is clear that the marginal variance of the states is dependent on the values of the $A$ and $\boldsymbol{\Sigma}$, which in turn complicates estimation as the marginal variance of the states must be constant throughout estimation else identification issues will arise. Unlike with cross-sectional modeling, the marginal variance of the states cannot be fixed by setting values in one of $\boldsymbol{\Sigma}$ or $\mathbf{A}$ to a constant, as the marginal variance of the states will still depend on the values of the other matrix. To account for this, we use the following set of dynamic constraints. First, we assume that $\boldsymbol{\Gamma}$ and $\boldsymbol{\Sigma}$ have the following forms.

\begin{equation}
\boldsymbol{\Gamma} =
\begin{bmatrix}
1      & \gamma_{12} & \gamma_{13} & \cdots & \gamma_{1N} \\
\gamma_{12} & 1      & \gamma_{23} & \cdots & \gamma_{2N} \\
\gamma_{13} & \gamma_{23} & 1      & \cdots & \gamma_{3N} \\
\vdots & \vdots & \vdots & \ddots & \vdots \\
\gamma_{1N} & \gamma_{2N} & \gamma_{3N} & \cdots & 1
\end{bmatrix}
\end{equation}

\begin{equation}
\boldsymbol{\Sigma} =
\begin{bmatrix}
\sigma_{1} & 0          & 0          & \cdots & 0          \\
0          & \sigma_{2} & 0          & \cdots & 0          \\
0          & 0          & \sigma_{3} & \cdots & 0          \\
\vdots     & \vdots     & \vdots     & \ddots & \vdots     \\
0          & 0          & 0          & \cdots & \sigma_{N}
\end{bmatrix}
\end{equation}
We assume that $\mathbf{A}$ results in a stationary process, which ensures that the marginal variance of $\mathbf{x}_t$ exists. These assumptions correspond to setting the scale of the states to 1, while allowing for arbitrary between state marginal correlations. The assumption regarding $\boldsymbol{\Sigma}$ correspond to the 
state innovation processes being independent of one another, above and beyond the state dynamics described by $\mathbf{A}.$

From here, given permissible values for $\mathbf{A}$, one can solve first for the off diagonal elements of $\boldsymbol{\Gamma}$ then by solving for the diagonal elements of $\boldsymbol{\Sigma}$. During estimation, the diagonal elements of $\boldsymbol{\Sigma}$ are updated whenever $\mathbf{A}$ is updated, resulting in the diagonal values of $\boldsymbol{\Sigma}$ being purely a function of $\mathbf{A}$.

While this identification technique is computationally expensive, as it requires solving systems of equations whenever $\mathbf{A}$ is changed, it has a significant advantage: because the model is identified without fixing values in the measurement model, estimated values of $\mathbf{A}$ are directly comparable for any measurement model. For example, we use this property later in this manuscript to directly compare $\mathbf{A}$ estimates when the measurement model is the correct GR model vs. a linear approximation. 

\subsection{State Estimation Via Particle Filtering}

There are two halves to the estimation of state-space models. The first half is the estimation of the state values themselves, given a set of parameter estimates and the observed data. Estimating the expected value of a state at time $t$ (i.e., $\mathbf{x}_t$) given the parameter values and observed variables up to and including time $t$ is known as \textit{filtering}. In state-space modeling, many different approaches to filtering have been developed. One set of approaches can be classified as analytic filtering, and includes the classical Kalman filter, as well as many of its newer variants. Unfortunately, analytic filtering approaches rely on the filtering distribution (i.e. $\mathbf{x}_t|\mathbf{y}_{1:t}, \boldsymbol{\theta}$ with $\boldsymbol{\theta}$ standing for the entire parameter set) being analytically tractable. In the case of our ordinal outcome state-space models, the filtering distribution is not analytically tractable, and a different approach is needed.

The issue of intractable latent variable distributions in cross-sectional item response theory models is well known, and has been solved by either using numeric integration methods or Bayesian estimation approaches \citep{baker_item_2004}. Here, we use \textit{particle filtering} \citep{moral_particle_2014}, an estimation method conceptually related to Bayesian MCMC, which we describe generally in Algorithm \ref{alg:pf}.

\begin{algorithm}
\caption{Basic Particle Filter for State Estimation}\label{alg:pf}
\begin{algorithmic}
\State Initialize $K$ particles 
$\mathbf{x}^{k}_0 \sim f_0$
\For{t in 1:T}
\State Sample $\mathbf{x}^{k}_t \sim f(\mathbf{x}^{k}_{t-1}|\boldsymbol{\theta})$
\State Calculate $w_k = p(\mathbf{y}_t|\mathbf{x}^{k}_t,\boldsymbol{\theta})$
\State Normalize $w_k: w^*_k = w_k/\sum_k w_k$ 
\State Sample with replacement $K$ particles from $\mathbf{x}^{k}_t$ with probability $w^*_k$ with replacement. 
\EndFor
\end{algorithmic}
\end{algorithm}

A particle filter is a simulation-based state estimation method that iteratively takes 
informed guesses at the values of $\mathbf{x}_t$ using the state dynamics model $f(\mathbf{x}_t|x_{t-1}, \boldsymbol{\theta})$, determines how good those guesses were by evaluating the likelihood of the observed data given the particles ($w_k = p(\mathbf{y}_t|\mathbf{x}^{k}_t,\boldsymbol{\theta})$), and then selects the best performing particles at each timepoint via resampling with replacement. While computationally expensive, particle filtering is a turnkey solution to the filtering problem, requiring only the state dynamics and measurement distributions, while making no assumptions about the form of them. In the case of the GR state-space models of Eqs 5,6, and 7, $f(\mathbf{x}_t|\mathbf{x}_{t-1}, \boldsymbol{\theta})$ is Eq 5, while $p(\mathbf{y}_t|\mathbf{x}^{i}_t,\boldsymbol{\theta})$ is Eq 8. The distribution of the initial values $\mathbf{x}_0$ is $f_0 = \text{MVN}(\mathbf{0}, \mathbf{I})$. Finally, $\mathbb{E}[\mathbf{x}_t|\mathbf{x}_{t-1}, \mathbf{y}_t, \boldsymbol{\theta}]$ is simply $\sum^{K} \bar{w}_k\mathbf{x}^{k}_t$.

\subsection{Parameter and State Estimation via the MIF2 approach of \citet{ionides_inference_2015}}

While the particle filter described by Algorithm \ref{alg:pf} can be used to estimate the expected value of the states, careful readers will notice that the model parameters $\boldsymbol{\theta}$ are assumed to be fixed and known. While this is not an issue in common engineering applications of state-space models where the model parameters are known, in the analysis of psychological or behavioral data the estimation of the parameters is of greatest importance. To tackle this side of the estimation problem, we employ yet more particle filtering.

To simultaneously estimate parameters and states we use the Multiple Iterated Filtering approach of \citet[MIF2;][]{ionides_inference_2015}. This algorithm is a general estimation routine for partially observed Markov processes (i.e., most state-space models), and can be conceptualized as a particle filter combined with with an optimization technique called simulated annealing. Simulated annealing is a method to find the global optima of a surface by sequentially perturbing an estimate and updating the estimate when the perturbation results in a higher value. Over the course of the optimization, how much the estimate is perturbed is reduced to 0, resulting in the final estimate that has ``frozen'' at the global optima \citep{kirkpatrick_optimization_1983}. The authors of \cite{ionides_inference_2015} have developed an optimized implementation of the MIF2 algorithm within their \texttt{pomp} R package \citep{king_pomp_2023,king_statistical_2016}, and our implementation of the GR model makes use of it. One key advantage of the MIF2 algorithm is that it does not require any analytic derivations of gradients or hessians, and only requires the ability to simulate from implied state distributions and evaluate the log-likelihood of the data given a set of parameter estimates. This makes the approach able to work with arbitrary measurement and state distributions.

The MIF2 algorithm applies particle filtering to both the states and parameter values, sampling, evaluating and resampling both over many iterations. The state filtering component of the algorithm is similar to what is described in Algorithm \ref{alg:pf}, while the parameter filtering component makes use of a sampling distribution that progressively reduces its variance over the iterations. This causes the estimation routine to first search the parameter space, then freeze in an area of high likelihood, ideally the MLE. This converging behavior would not occur if the MIF2 algorithm used a set perturbation variance \citep{ionides_inference_2015}.The MIF2 algorithm is described in Algorithm \ref{alg:mif2}.

\begin{algorithm}[H]
\caption{MIF2 Algorithm of \citet{ionides_inference_2015}}\label{alg:mif2}
\begin{algorithmic}
\State Initialize $K$ particles ${\boldsymbol{\theta}^{0,k}_0 = \boldsymbol{\theta}^*}$
\For{m in 1:M}
\State Sample $\boldsymbol{\theta}^{m,k}_0 \sim \text{MVN}(\boldsymbol{\theta}^{m-1,k}, c^m\boldsymbol{\Theta})$ for $k \in 1:K$
\State Sample $\mathbf{x}^{m,k}_0 \sim f_0(\boldsymbol{\theta}^{m,k})$ for $k \in 1:K$
\For{t in 1:T}
\State Sample $\boldsymbol{\theta}^{m,k}_t \sim \text{MVN}(\boldsymbol{\theta}^{m,k}_{t-1}, c^m\boldsymbol{\Theta})$ for $k \in 1:K$
\State Sample $\mathbf{x}^{m,k}_t \sim f(\mathbf{x}^{m,k}_t|\mathbf{x}^{m,k}_{t-1},  \boldsymbol{\theta}^{m,k}_t)$ for $k \in 1:K$
\State Calculate $w_k = p(\mathbf{y}_t|\mathbf{x}^{m,k}_t,\boldsymbol{\theta}^{m,k}_t)$  for $k \in 1:K$
\State Normalize $w_k: w^*_k = w_k/\sum_k w_k$ 
\State Sample with replacement $K$ particles from $\{\boldsymbol{\theta}^{m,k}_t,\mathbf{x}^{m,k}_t\}$ with probability $w^*_k$. 
\EndFor
\EndFor
\end{algorithmic}
\end{algorithm}
In Algorithm 2, $\boldsymbol{\Theta}$ is a diagonal covariance matrix with analyst specified variances, and $c^m$ is a positive and decreasing sequence such that $c^m \rightarrow 0 $ as $m \rightarrow M$. \citet{ionides_inference_2015} implement this as a geometrically decreasing sequence. The tuning values for this algorithm are the number of particles $K$, the number of iterations $M$, the parameter perturbation covariance matrix $\boldsymbol{\Theta}$ and the cooling sequence $c^m$. 
\citet{ionides_inference_2015} show that the MIF2 algorithm converges to the MLE under mild regularity conditions, making this approach a general estimating technique for state-space models that does not rely on tractable analytics or numerical derivatives. Now we can map the terms in the MIF2 algorithm to the specifics of the GR state-space model, which are contained in Table \ref{tab:grmif2}.

\begin{table}[H]
\centering
\caption{GR State-Space to MIF2 Mapping}
\label{tab:grmif2}
\begin{tabular}{cc}
\hline
MIF2 Terms                                                & GR State-Space Terms                                          \\ \hline
$\boldsymbol{\theta}$                                     & $\mathbf{A}$, $\{\beta_{ij} \text{ } \forall \text{ } i,j \}$\\
$f_0(\boldsymbol{\theta}^{m,k})$                          & $\text{MVN}(\mathbf{0}, \mathbf{I})$                                 \\
$f(\mathbf{x}^{m,k}_t|\mathbf{x}^{m,k}_{t-1}, \boldsymbol{\theta}^{m,k}_t)$ & $\text{MVN}(A^{m,k}_t \mathbf{x}^{m,k}_{t-1}, \mathbf{\Sigma_t^{m,k}})$         \\
$p(\mathbf{y}_t|\mathbf{x}^{m,k}_t,\boldsymbol{\theta}^{m,k}_t)$ &
  \begin{tabular}{c}
  $p(\mathbf{y}_{it} = j | \mathbf{x}^{m, k}_t) = p(\mathbf{y}_{it} > j | \mathbf{x}^{m,k}_t) - p(\mathbf{y}_{it} > j-1 | \mathbf{x}^{m,k}_t)$ \\
  $p(\mathbf{y}_{it} > j | \mathbf{x}^{m, k}_t) = \frac{1}{1+\exp[-\alpha_i (\delta_i(\mathbf{x}^{m,k}_t) - \beta_{ij}^{m,k})]}$
  \end{tabular} \\ \hline
\end{tabular}
\end{table}

We implement our GR state-space models in the \texttt{genss} R package \citep{falk_netlabuvagenss_2023}. This package is under active development, and will be expanded with additional measurement models and state dynamics in the near future.

Note that the discrimination parameters $\alpha_i$ are not estimated in this set, and are instead assumed to be fixed and known. In this manuscript, we fixed $\alpha_i = 1$ (i.e., a Rasch model version of the Graded Response Model) as the MIF2 algorithm had considerable difficulties estimating $\alpha_i$ freely. This is not due to model identification issues, but rather due to the nature of $\alpha_i$. Specifically, the relative magnitude of $\alpha_i$ only makes a difference to the likelihood of the data when close to $0$, while for example $\alpha_i = 5$ and $\alpha_i = 10$ would result in nearly identical likelihood surfaces. This appears to cause difficulty with MIF2's exploration of the likelihood surface. This is a limitation of the estimating approach, and we discuss possible solutions in the Discussion section.

\subsection{Standard Errors via Slice Likelihoods}

While the MIF2 algorithm converges to the MLE under mild regularity conditions, it does not produce the information necessary to estimate the standard errors of the parameters. Calculating standard errors requires some knowledge of the curvature of the likelihood surface at the MLE, and as MIF2 does not use gradient information, the Fisher information matrix needs to be approximated. One potential way of doing this is to calculate numerical derivatives, however this is complicated by the stochastic nature of particle filtering. Here, we take the slice likelihood approach described by \citet{ionides_inference_2006} to approximate standard errors.

Slice likelihood relies on the \textit{local asymptotic normality} \citep[LAN;][]{le_cam_locally_1990} property of the likelihood surface near the MLE. This property says that the likelihood surface is approximately multivariate normal in the area around the MLE. This in turn allows one to estimate the curvature of the likelihood surface at the MLE using a series of quadratic regressions, which can then be used to approximate the Fisher information matrix. Slice likelihood involves evaluating the likelihood at a series of slices surrounding the MLE, usually along each parameter (hence, ``slicing'' along each parameter axis). Those values are then used in the series of quadratic regressions to estimate the curvature of the likelihood surface. The details of the implementation can be found in the Supplementary Materials of \citet{ionides_inference_2006}, and we include our implementation in the \texttt{genss} package \citep{falk_netlabuvagenss_2023}. These standard error estimates are approximations and are known to be more liberal (i.e. smaller) than the true standard errors. We evaluate the performance of these slice likelihood standard errors in a simulation study below. Other methods for calculating standard errors when using particle filtering methods for estimation is an active area of research, and we discuss other potential options in our Discussion section.

\section{A Simulation Study}

With the estimation details of the GR state-space model established, we now describe a simulation study that serves two purposes: 1) to evaluate how MIF2 and slice-likelihood standard errors serve in fitting GR state-space models to data generated from a GR state-space model, and 2) to evaluate how a normal measurement state-space model (the \textit{linear approximation} approach) performs when fit to data generated from a GR state-space model. 

\section{Methods}

\subsection{Data Generating Model}

In all conditions, the data generating model was a GR state-space model with 2 states and a varying number of indicators per state. The number of states was fixed at two to avoid radically increasing the number of conditions and computational cost of fitting the data, while still being able to demonstrate the estimation of cross-regressive effects. The form of the $\mathbf{A}$ matrix is $\begin{bmatrix}
    AR & CR \\
    0  & AR \\
\end{bmatrix}$ with $AR$ referring to the autoregressive effects, and $CR$ referring to the cross-regressive effect of the first state on the second state. The $\boldsymbol{\Sigma}$ used to generate the data is determined in accordance to the identification constraints described previously. In all cases, each observed GR variable is associated with a single state, which we assume is known when we fit the models to the simulated data. The $\alpha_i$ parameters are set to 1 and are considered fixed and known during estimation. The number of items per state varies by condition, but is the same for each of the two states. Finally, $\beta_{ij}$ values were determined in two ways. In the \textit{equal} GRM threshold condition, $\beta_{ij}$ were calculated by centering the sequence $[1,2,\dots, j]$ on $0$, so that, for example, a item with 7 response categories would have $\beta_{ij} = [-2.5,-1.5,-.5, .5, 1.5, 2.5]$. In this \textit{equal} condition, the $\beta_{ij}$ sets for each observed variable are the same, corresponding to the situation where each observed variable has identical measurement properties. In the \textit{offset} GRM threshold condition, the $\beta_{ij}$ sets for each item within a given state were offset from one another by either the number of items divided by the number of $\beta_{ij}$ per item, if that offset is less than 1.25, and 1.25 if it was calculated to be greater than 1.25. In the case of three items per state, with 7 response options per item, this results in the following set of $\beta_{ij}$, $\beta_{1j} = [-3,-2,-1, 0, 1, 2], \beta_{2j} = [-2.5,-1.5,-.5, .5, 1.5, 2.5], \beta_{3j} = [-2,-1,0,1,2,3]$. This condition corresponds with more typical measurement in psychometric scales, where the items were designed to cover a range of latent values. 

\begin{table}[H]
\centering
\caption{Simulation Factors}
\label{tab:simconds}
\begin{tabular}{ll}
\hline
Simulation Factor            & Values        \\ \hline
Number of Timepoints         & 100, 500      \\
Number of Items per State    & 3, 6          \\
Number of Responses per Item & 3, 7          \\
AR Parameter Value           & .3, .7        \\
CR Parameter Value           & 0, .25        \\
GRM Thresholds               & Equal, Offset \\ \hline
\end{tabular}
\end{table}

Table \ref{tab:simconds} lists the simulation factors and values. All factors were fully crossed, which resulted in 64 conditions. 300 iterations per condition were generated.

\subsection{Analysis Models}

All generated datasets were fit with a GR model and a linear approximation model. For both the GR model and linear approximation model, $\mathbf{A}$ was unconstrained and $\mathbf{\Sigma}$ was calculated using the identification constraint described above. Recall that this allows for direct comparison of the estimated values of $\mathbf{A}$ between the GR and linear approximation models. In both cases, the initial values for $\mathbf{A} = \begin{bmatrix}
    .1 & 0 \\
    0  & .1 \\
\end{bmatrix} $. In the fit GR models, all $\beta_{ij}$ are freely estimated. The initial values for $\beta_{i1}$ were set to -2, and the initial values for the intervals between subsequent $\beta_{ij}$'s was set to $.36$.

In the linear approximation models, the matrix $\mathbf{C}$ from Eq 2. was constrained to freely estimate non-0 state-item loadings, corresponding to a known measurement structure with unknown parameter values. The initial values for non-zero loadings were set to 1. Initial values for diagonal elements of $\boldsymbol{\Psi}$ were set to 1.

Finally, 4 separate runs of the MIF2 algorithm were performed on for each model as per \citet{ionides_inference_2015} and the estimated coefficients were averaged into a final estimate. Slice likelihood standard errors were calculated based on the averaged estimates. In all cases, 1000 particles were used with 250 MIF2 iterations. The cooling proportion per 50 iterations was set to .05 (i.e. after 50 iterations, the variance of the parameter perturbation distribution, $\boldsymbol{\Theta}$ is .05 of the starting values). For all parameters the initial perturbation standard deviation was set to $.3$.

\subsection{Outcomes}

As our interests were chiefly in evaluating the GR models performance in recovering state dynamics and comparing its performance against the linear approximation model, our outcomes focus on the $\mathbf{A}$ matrix. While $\beta_{ij}$ parameters were estimated, we do not present outcomes related to them in this manuscript due to space concerns. 

We present and discuss the following outcomes:

\begin{itemize}
    \item \textbf{True-Estimated State Correlation} - We operationalize this outcome as the Spearman's correlation between the true state values, and the expected values of the states calculated via MIF2. This is a general measure of model performance, with values close to 1 indicating perfect recovery of the unobserved states, and values away from 1 indicating poor recovery.
    \item \textbf{Relative Bias of the Autoregressive Parameters} - This signed relative bias evaluates how well the models recover the autoregressive parameters. Values close to 0 indicate unbiased recovery, while non-zero values indicate bias.
    \item \textbf{Bias of the Cross-Regressive Parameter} - As one of our CR conditions set the parameter to 0, to examine the bias of the CR parameter we use bias rather than relative bias. 
    \item \textbf{95\% Wald Type Confidence Interval Coverage} - To evaluate how accurate our slice likelihood standard errors were, we calculated a simple 95\% confidence interval (i.e. $\theta \pm 1.96 \times SE$) and determined the coverage. This was calculated for both the AR and CR parameters.
\end{itemize}

Additionally, we present plots and tables of slice standard errors in the Supplementary Materials.

\section{Results}

\subsection{True-Estimated State Correlations}

\begin{figure}[H]
    \centering
    \includegraphics[width=.99\textwidth]{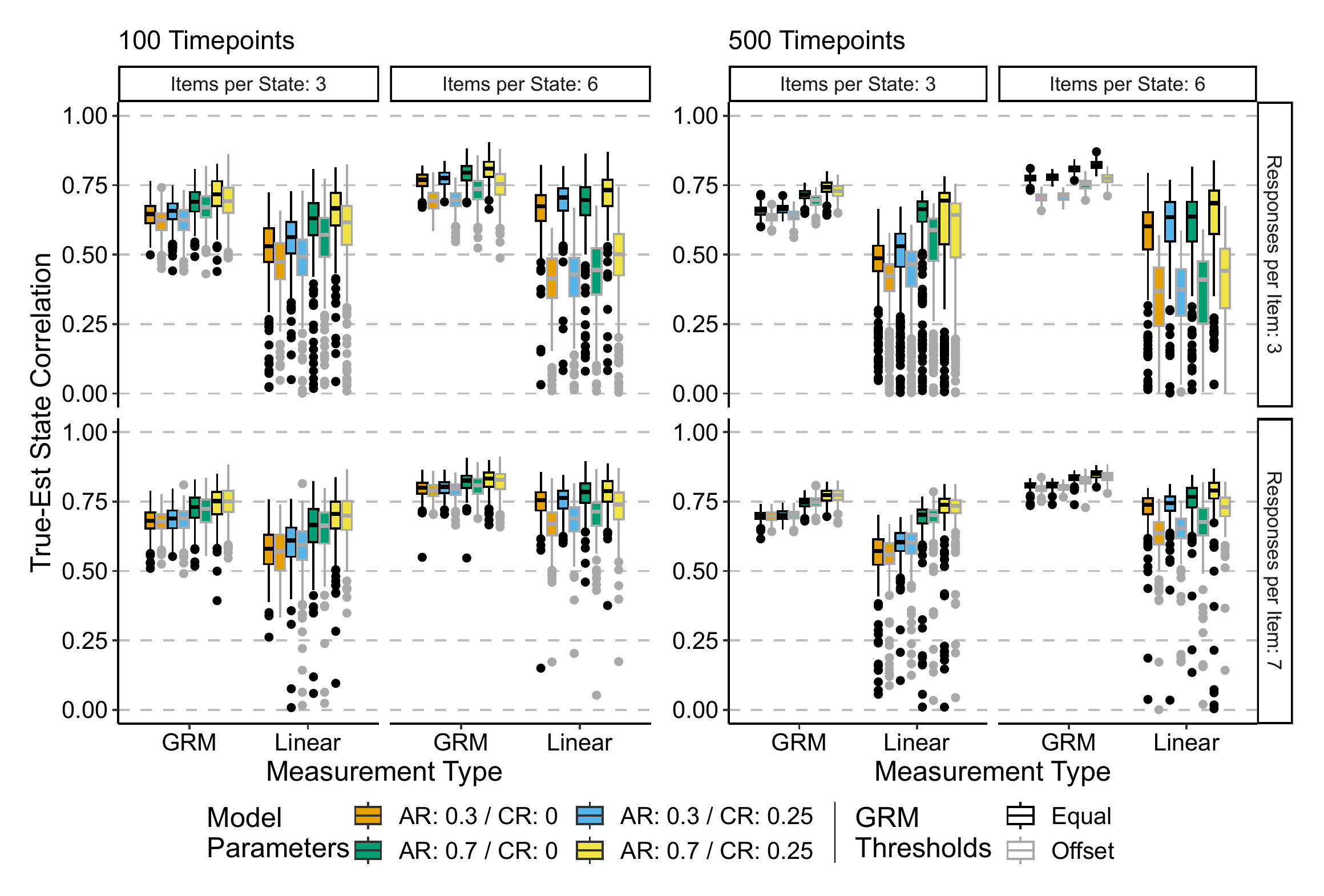}
    \caption{True - Estimated State Spearman's Correlations. Central black line on boxplot denotes median, box denotes $25\%-75\%$ interquartile range, whiskers denote $1.5 \times IQR$. AR refers to the autoregressive parameter while CR refers to the cross-regressive parameter.}
    \label{fig:staterecov}
\end{figure}

Figure \ref{fig:staterecov} shows the true-estimated state correlations across all conditions. Median values per condition are contained in Supplementary Materials Tables S1 and S2. There are several notable findings here. First, in every condition the GR state-space model resulted in better overall recovery of the state values, with the maximum median recovery of $\rho_s = .849$  occurring in the 500 timepoint, AR: 0.7 / CR: 0.25, equal thresholds, 6 items per state with 7 response categories. That the best recovery of the state values occurs in this condition is no surprise, given that it is the most generous data-quality wise. The median recovery for the linear approximation model for the same condition was $\rho_s = .749$. The worst recoveries for the GRM model were in the 100 timepoint, AR: 0.3 / CR: 0 or .25, offset threshold, 3 items per state, 3 response categories per item condition, with a median recovery of $\rho_s = .625$. The linear approximation model had median recoveries of $\rho_s = .474 \text{ and } .491$. In addition to the GR models performing better than the linear approximation models with respect to overall level of state recovery, the GR models also had less variable recoveries across simulation iterations than the linear approximation models. The linear approximation models showed substantial negative skew in the true-estimated state correlations, suggesting that the linear approximation model struggled to recover the state values more often than the GR model.

In terms of the simulation factors effects on the state recovery, both GR and linear approximation models had the same general pattern of effects, with the notable exception of the effect of number of timepoints. In the case of the GR model, increasing the number of timepoints from 100 to 500 uniformly decreased the variability of the true-estimated state correlations within a condition, and increased the overall true-estimated state recovery. However, for the linear approximation model in the 3 response categories per item conditions (top row of Fig \ref{fig:staterecov}), increasing the number of timepoints from 100 to 500 led to an increase in variability and a overall slight decrease in the median level of recovery. This was not the case for the linear approximation models in the 7 response categories per item conditions (bottom row of Fig \ref{fig:staterecov}), though both the reduction in variability and improvement in median recovery was modest compared to that of the GR model. 

Increasing the number of items per state generally improved the recovery of the state values, as did increasing the number of response categories per state. Increasing the magnitude of the AR and/or CR parameters also generally increased the median recovery across all other conditions. Interestingly, having the GR threshold parameters (i.e. $\beta_{ij}$) be equal across items led to better overall recovery than having the GR threshold parameters be offset. Additionally, there appears to be an interaction between the offset of the GR thresholds and the number of items per state, where the reduction in recovery when GR threshold parameters were offset is greater with more items per state. This result is most noticeable in the linear approximation model, but occurs with lower magnitude in the GR models as well.

\subsection{Relative Bias of the AR Parameter}

\begin{figure}[H]
    \centering
    \includegraphics[width=.99\textwidth]{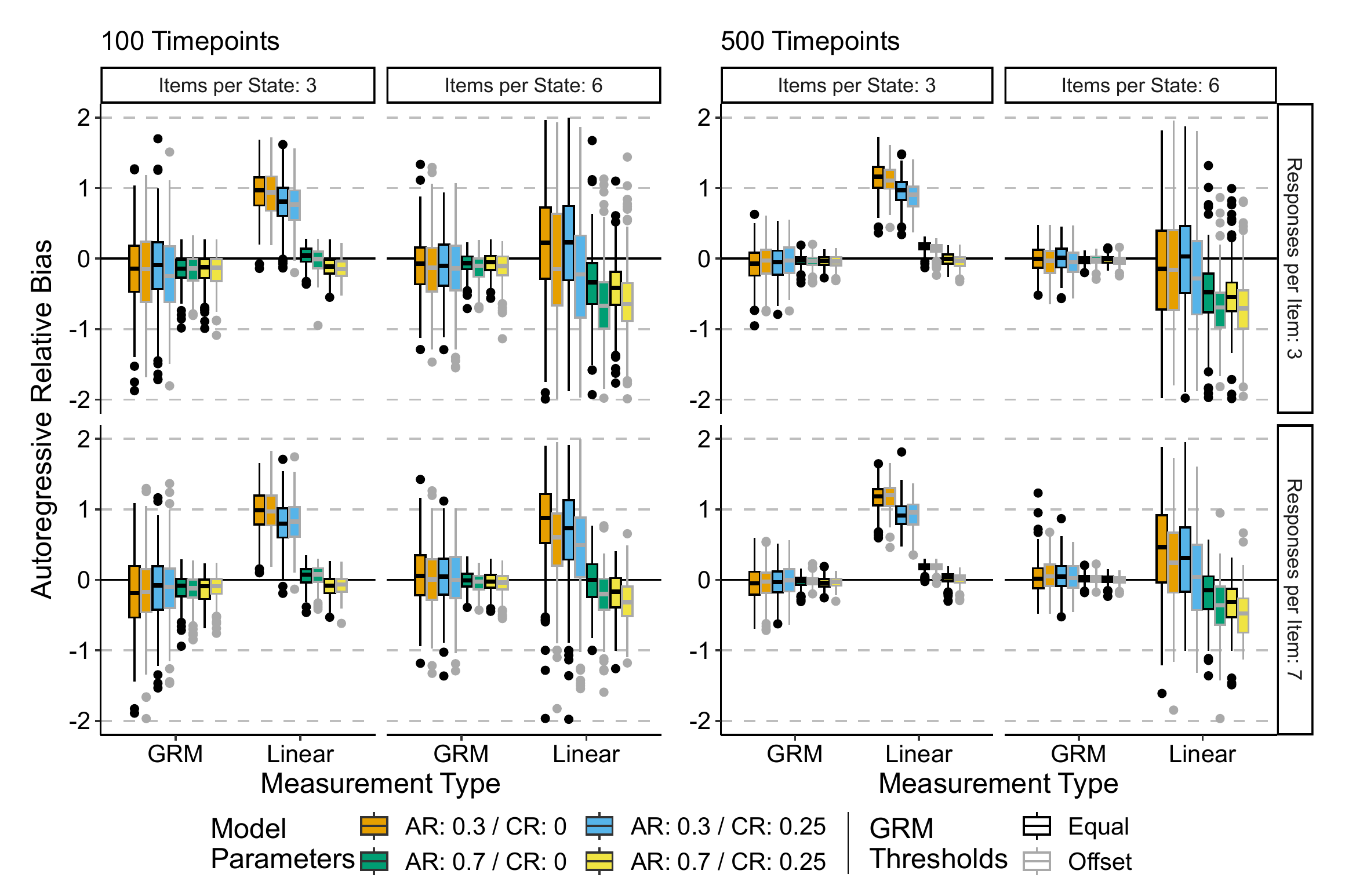}
    \caption{Autoregressive Parameter Relative Bias. Central black line on boxplot denotes median, box denotes $25\%-75\%$ interquartile range, whiskers denote $1.5 \times IQR$. AR refers to the autoregressive parameter while CR refers to the cross-regressive parameter.}
    \label{fig:arrelbias}
\end{figure}

Figure \ref{fig:arrelbias} visualizes the relative bias of the AR parameters. Table S1 and S2 in the Supplementary Materials contain median relative bias values across all conditions. 

First, the results shown in Figure \ref{fig:arrelbias} suggest that the GR model results in asymptotically unbiased estimates of the AR coefficients. Across all conditions, the median relative bias is close to 0, and the variability of this bias decreases with increased numbers of timepoints. As one would expect, larger magnitudes of the AR coefficient result in lower relative bias.

However, the linear approximation model did not result in unbiased estimates of the AR parameters. Several patterns of findings stand out. First, when the AR parameters were small ($AR = 0.3$), the linear approximation model resulted in AR parameters that were positively biased, with the greatest median relative bias of approximately .75 to 1.25 occurring in the 3 items per state conditions. In the 6 items per state conditions, the variability of the relative bias was radically increased relative to the variability in the 3 items per state condition. For stronger AR effects ($AR = .7$), the linear approximation models resulted in relatively unbiased estimates only during in the 3 items per state conditions, while for the 6 items per state conditions, there was a consistent negative bias for the AR coefficients. Notably, the variability of the bias was minimally impacted by increasing the number of timepoints from 100 to 500.

\subsection{Bias of the CR Parameter}

\begin{figure}[H]
    \centering
    \includegraphics[width=.99\textwidth]{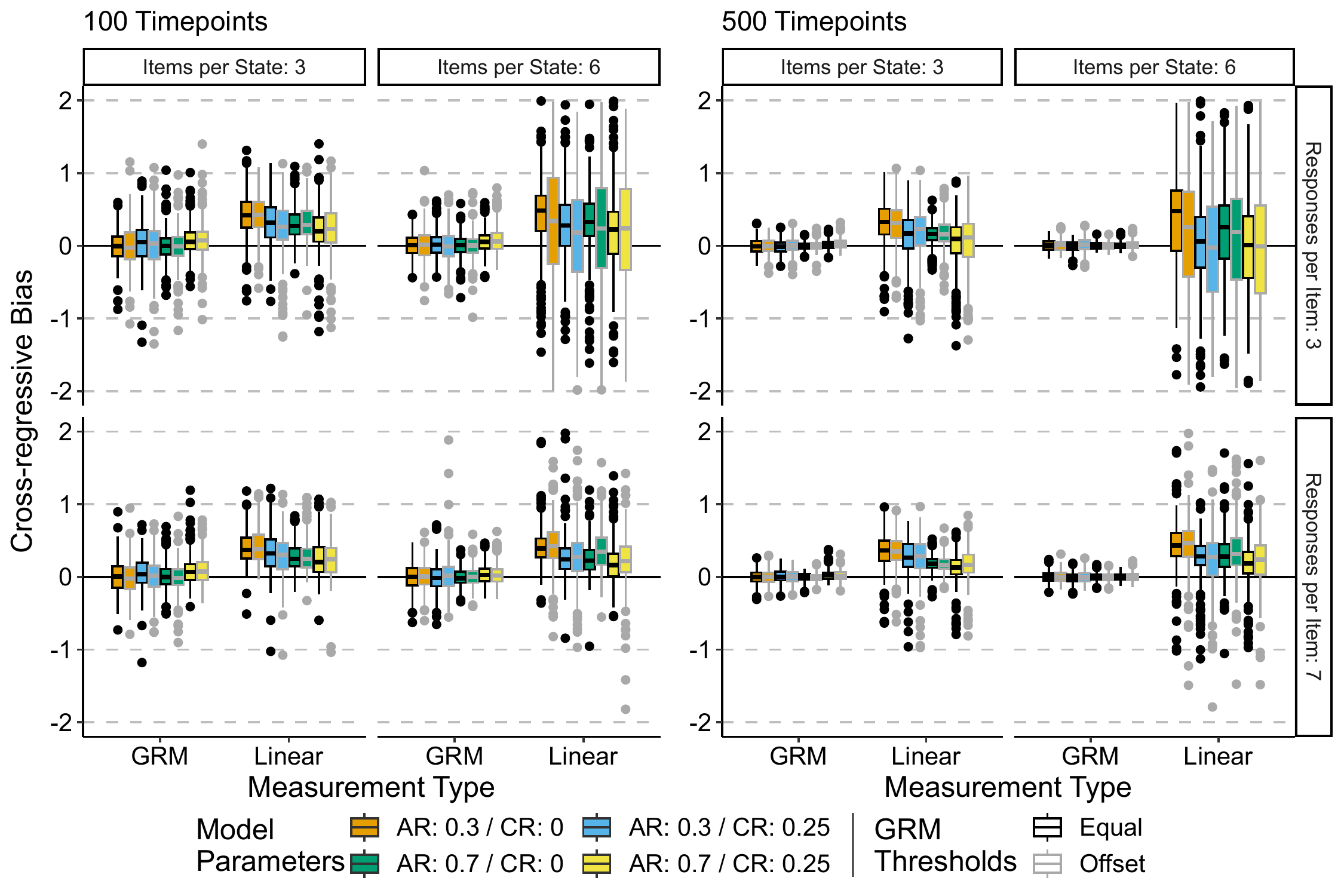}
    \caption{Cross-regressive Parameter Bias. Central black line on boxplot denotes median, box denotes $25\%-75\%$ interquartile range, whiskers denote $1.5 \times IQR$. AR refers to the autoregressive parameter while CR refers to the cross-regressive parameter. Note that this figure shows bias rather than relative bias.}
    \label{fig:crbias}
\end{figure}

Figure \ref{fig:crbias} shows the bias of the CR parameter across all conditions. Tables S3 and S4 in the Supplementary Materials contain the median values.

The results shown in Figure \ref{fig:crbias} suggest that the GR model results in unbiased estimates of the CR parameter, with variance decreasing and median value becoming more 0 centered with an increase in the number of time points. For the linear approximation model, the bias of the CR parameter is consistently positive, and this bias does not improve with increased sample sizes. Consistent with the AR parameter, the 6 items per state/3 response categories per state conditions exhibited increased variability in the bias for the linear approximation models relative to the 3 items per state conditions. It appears that increasing the magnitude of the CR parameter does slightly improve bias, however this difference is slight. Finally, there is a less pronounced effect of the GRM thresholds being equal vs. offset. The offset conditions appeared to result in slightly higher variance in the bias for the linear approximation models, but there is little difference in the median bias.

\subsection{Parameter Coverage}

\begin{figure}[H]
    \centering
    \includegraphics[width=.99\textwidth]{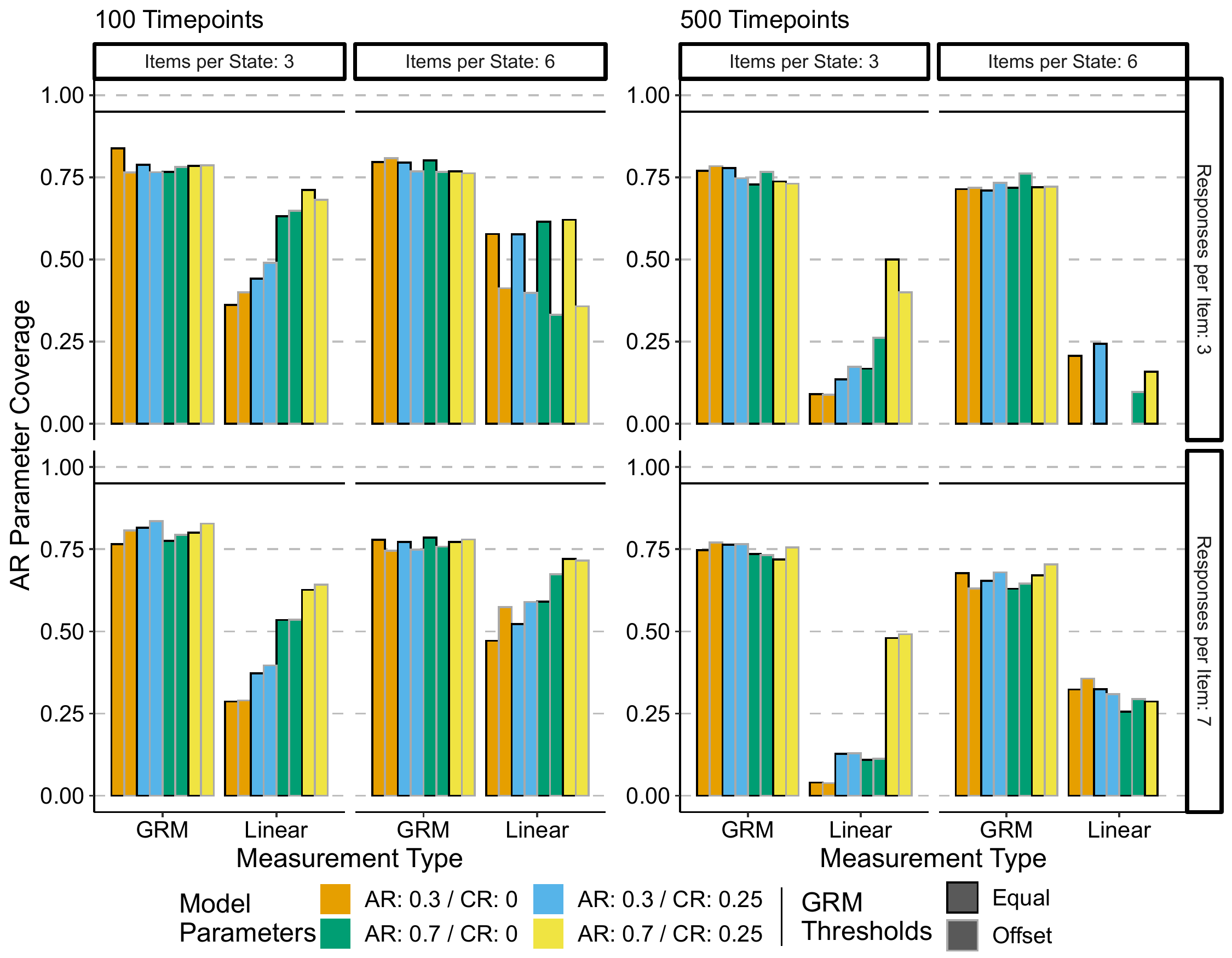}
    \caption{Autoregressive Parameter Coverage. Height of bars represent coverage.}
    \label{fig:arcov}
\end{figure}

Figure \ref{fig:arcov} shows the coverage of the 95\% Wald type confidence interval using the slice likelihood SEs. Tables S1 and S2 in the Supplementary Materials lists the coverage values per condition. 

First, we note that across most conditions the coverage of the AR parameter in the GR models was consistently close to 75\%, with the exception of the 500 timepoints, 6 items per state, 7 responses categories per state conditions, where the coverage was approximately 70\%. There were minimal differences between coverage with respect to parameter values or GRM thresholds for the GR model. This is consistent with the GR models resulting in unbiased estimates of the parameters, with the lower than nominal coverage consistent with the slice-likelihood SE being a liberal estimate. 

On the other hand, the coverage of the AR parameter when using the linear approximation model was less than ideal. The coverage was not consistent across conditions, with stronger magnitude parameters having better coverage overall than smaller parameter conditions. These results are consistent with the patterns of bias shown by the linear approximation models, combined with the liberal nature of the slice likelihood standard errors.

\begin{figure}[H]
    \centering
    \includegraphics[width=.99\textwidth]{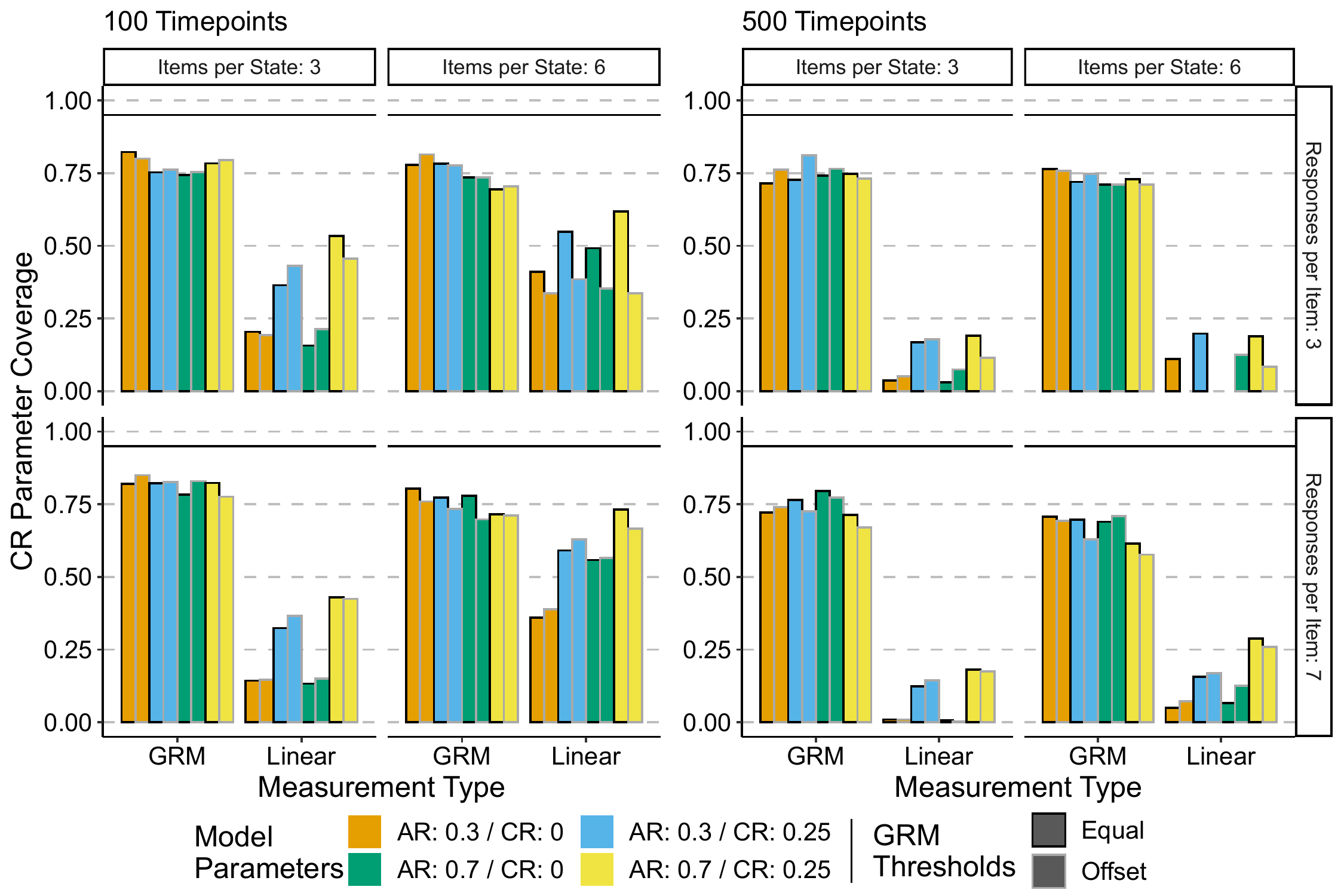}
    \caption{Cross-regressive Parameter Coverage. Height of bars represent coverage.}
    \label{fig:crcov}
\end{figure}

Figure \ref{fig:crcov} shows the coverage of the CR parameter over all conditions. Tables S3 and S4 in the Supplementary Materials contain the values of coverage.

The coverage results for the CR parameter are consistent with the coverage results for the AR parameter. Overall the coverage of the GR models was approximately .75\% with the exception of the 500 timepoint, 6 items per state, 7 responses per item conditions, where the coverage values range from approximately 55\% to 70\%. Again, this is consistent with unbiased parameter estimates combined with overly liberal standard errors. 

Compared to their coverage of the AR parameter, the linear approximation models have considerably worse coverage of the CR parameter. Coverage is best when the CR parameter is non-zero, however coverage decreases with increased sample size. This is consistent with the bias estimates that the linear approximation models provide.

\subsection{Slice Likelihood Standard Errors}

Figures S1 and S2 in the Supplementary Materials, as well as Tables S1, S2, S3, and S4 display and contain the slice likelihood standard error estimates. We briefly describe the overall pattern of findings here.

For both the AR and CR coefficient, two interesting patterns emerge. In the 3 items per state conditions, the median slice likelihood SEs of the linear approximation models are slightly less than that of the GR models, though this difference becomes neglible at 500 timepoints. Conversely, in the 6 items per state conditions, the linear approximation models have greater median slice likelihood SEs and more variability in the distribution of SEs than the GR model. This is reduced but not eliminated at 500 timepoints.

\section{Discussion}

In this manuscript, we present an estimation strategy for state-space models with graded response measurements, suitable for use with ordinal measurements such as Likert scales. We demonstrate that the estimation routine results in unbiased parameter estimates, and compare the performance of our model with that of a linear approximation approach, consistent with fitting a dynamic structural equation model (in an $N = 1$ setting) to ordinal measurements. The findings of our simulation study suggest that the linear approximation approach results in biased estimates of the state dynamics, and should be used with considerable caution.

While there are a number of methods for estimating state-space type models with ordinal outcomes, such as Bayesian approaches in DSEM \citep{asparouhov_dynamic_2018} or the work of van Rijn \citep{van_rijn_categorical_2008} using generalized linear modeling, the estimation approach we describe here is innovative in a number of ways. First, the identification constraint we develop here involves only the parameters for state dynamics, allowing comparable estimates to be obtained from models with differing measurement models. Second, the use of particle filtering, and more generally the MIF2 \cite{ionides_inference_2015} algorithm, provides a general estimator that does not require the use of gradients that is suitable for any measurement model and/or any mix of measurement models, while both DSEM and the approach of van Rijn are restricted to specific classes of outcomes. While the estimation technique itself relies on simulation akin to standard methods in Bayesian estimation, the estimation is developed in a frequentist framework, requires less computation than full Bayesian estimation, and we provide a means of calculating approximate standard errors using slice-likelihoods. We emphasize that the MIF2 algorithm is not specific to ordinal measurement outcomes, and can be used to estimate models with any form of measurement (and for that matter state dynamics). Finally, we provide an open source implementation of the GR and linear approximation models used in this manuscript in the R package \texttt{genss} \citep{falk_netlabuvagenss_2023}.

Our simulation study demonstrates that the GR model results in unbiased estimates of the state-dynamics and that the slice-likelihood standard errors result in consistent coverage across conditions. However, this coverage was not at the nominal level of 95\%, indicating that the slice-likelihood standard errors are overly liberal. There are a number of approaches for obtaining accurate SE estimates from models that are estimated using a particle filtering approach such as profile likelihoods \citet{ionides_inference_2015} or particle filter approximations of the Hessian \citep{spall_monte_2005,chada_unbiased_2022}. However, these approaches are orders of magnitude more computationally expensive than slice-likelihood, requiring in the case of the profile likelihoods re-estimation of thousands of models with slight variations in fixed parameter values. Refining the standard error estimates for these models is of paramount interest, as accurate SEs are necessary for accurate inference. That being said, the slice-likelihood standard errors are fairly consistent in their coverage in the GR model across conditions, suggesting that one can alleviate these issues by using more strict Wald type confidence intervals. For example, we found that a 99.8\% Wald CI (i.e. $\theta \pm 3.09 \times SE$) provided approximate 95\% coverage in the GR models. As the slice-likelihood SEs are more liberal, conservative statistical test thresholds should be used to help reduce the chance of false positive findings. We advocate for the use of 99.8\% confidence intervals, and that authors explicitly note the use of the slice-likelihood SEs along with their limitations. 

The linear approximation models performance in our simulation study revealed considerable issues with bias in both the AR and CR parameters, and that there was no set of conditions that completely alleviated the bias. Overall, it appeared that more response options per item (7 vs. 3), resulted in less variable estimates, but only minimally different estimates with respect to bias. This is somewhat consistent with the idea that an ordinal measurement with increasing number of response categories will converge to a continuous distribution (e.g. a 0-100 slider is more continuous than a 1-7 Likert scale), however the negligible improvement in bias suggests that the influence of increasing the number of response categories needs to be studied more. We also note that while increasing the number of timepoints improved bias for the GR model, this was not the case in the linear approximation models. Of course, this was not expected as increasing the sample size when fitting a mis-specified model does not make for less biased estimates, just more misplaced confidence in the biased estimates one obtains. Here, we suggest extreme caution is warranted in fitting continuous measurement state-space models to ordinal measurements, and strongly advocate for the use of an appropriate model such as the GR models presented here or other ordinal outcome state space models discussed previously.

There are a number of limitations to the modeling strategy presented here. First and foremost is the computational expense of the estimation. As the estimation routine is a simulation based method, and requires iterated particle filters, any model fit using MIF2 takes a considerable amount of time. In our simulations, it took approximately 15-20 minutes for a single model to fit, though that time heavily depends on the computational environment (15-20 minutes on a Linux desktop computer outfit for simulation studies, 20-30 minutes on a more typical Windows laptop.) There is also the limitation of needing to determine appropriate setting for the estimation itself (i.e. number of particles, cooling rate, number of iterations and the standard deviation of the perturbation distribution), along with the need to fit multiple runs of the same model to the data to determine if a global maximum has been obtained. These limitations can be ameliorated by optimizing the implementation, but should be taken into account when considering using this estimator. 

Another limitation of the approach we develop here is that the state-space models are $N = 1$, suitable for purely idiographic analysis. There are a couple of subtleties to this limitation that need to be discussed, as they are relevant for readers interested in fitting these models idiographically. First, because there is no sharing of information between multiple participant's models, both the measurement model and state dynamics will be specific to a given participant. However, state dynamics parameters can be compared between participants' models. Furthermore, as the expected values of the state distribution are assumed to be $0$ as per the identification constraints, the estimated state trajectories are not comparable between participants unless measurement parameters were assumed fixed and known.  If the $\beta_{ij}$ values were estimated per participant, the state trajectories are not directly comparable between participants, as the measurement parameters would  contain information as to the location of the participants' state values relative to the population. This limitation can be alleviated by using a calibrated set of items with known parameters, and modifying the GR state-space model to estimate participant specific mean vectors of states. If the same set of measurement parameters is used for all subjects in addition to constraining the marginal variance of the states to 1, then the state trajectories will be comparable between participants (and suitable for the calculation of percentiles relative to the population).

The limitations of the current model suggest a number of fruitful future directions. First, a means of analyzing multiple participants is needed. The estimation method is not suitable for analyzing multiple participants' data simultaneously in the same model due to computational expense, so we are investigating the use of meta-analytic techniques for analyzing the participant-specific  parameter estimates. This approach would preserve the idiographic nature of the models while allowing for an analysis of population heterogeneity/homogeneity. Second, as the MIF2 algorithm is a general estimator that allows for arbitrary measurement models, other measurement models such as distributions for counts, zero-inflated distributions, and other IRT measurement models, should be implemented and their performance evaluated. A low hanging fruit in this direction is that of zero-inflated distributions, which would improve our ability to model time series of rarer behaviors like substance use. Finally, the MIF2 algorithm itself should be studied to improve its speed. One core component of the MIF2 algorithm is the use of a symmetric perturbation distribution, which could be improved via integrating gradient information ala Hamiltonian MCMC \citep{neal_monte_1996} or by sharing information between particles ala a genetic optimization algorithm \citep{de_lima_particle_2011}. Finally, improved standard error estimates must be developed, potentially using the profile likelihood or particle filter estimation of the Hessian methods mentioned previously.

The analysis of intensive longitudinal data in psychology is only becoming more prominent, and researchers need analysis techniques that can appropriately account for the measurement properties of their data. We aim that the work presented in this manuscript provides a general estimation infrastructure for state-space modeling of ordinal psychological data, and look forward to expanding its capabilities in the near future.

\bibliography{References.bib}
\bibliographystyle{apacite}

\end{document}


\begin{titlepage}

\linespacing{1}
\title{Supplementary Materials for ``Ordinal Outcome State-Space Models for Intensive Longitudinal Data''}

\author{Teague R. Henry \\ Department of Psychology and School of Data Science, University of Virginia \\
  Lindley R. Slipetz \\ Department of Psychology, University of Virginia \\
  Ami Falk \\ Department of Psychology, University of Virginia\\
  Jiaxing Qiu \\ School of Data Science, University of Virginia\\
  Meng Chen \\ Health Sciences Center, University of Oklahoma}

\markboth{Psychometrika}{*}

\vspace{\fill}
\vspace{\fill}
\vspace{\fill}

\linespacing{1}\fontsize{8}{10}\selectfont

\end{titlepage}

\newpage
\linespacing{1}
\setcounter{table}{0}
\renewcommand{\thetable}{S\arabic{table}}
\section{Result Tables}

\begin{table}[H]
\centering
\caption{State Recovery and Autoregressive Parameter Outcomes for 100 Timepoint Conditions. N Items: Number of Items per State; N Responses: Number of Response Categories per Item.}
\label{tab:AR100tps}
\resizebox{\columnwidth}{!}{%
\begin{tabular}{lccccccccccc}
\hline
AR/CR Parameters &
  Spread &
  N Items &
  \multicolumn{1}{c|}{N Responses} &
  \multicolumn{2}{c|}{State Recovery} &
  \multicolumn{2}{c|}{Median Relative Bias} &
  \multicolumn{2}{c|}{Slice SE} &
  \multicolumn{2}{c}{Coverage} \\
 &
   &
   &
  \multicolumn{1}{c|}{} &
  GRM &
  \multicolumn{1}{c|}{Linear} &
  GRM &
  \multicolumn{1}{c|}{Linear} &
  GRM &
  \multicolumn{1}{c|}{Linear} &
  GRM &
  Linear \\ \hline
AR: 0.3 / CR: 0    & Equal  & 3 & 3 & 0.647 & 0.528 & -0.142 & 0.971  & 0.169 & 0.108 & 0.838 & 0.361 \\
AR: 0.3 / CR: 0    & Equal  & 3 & 7 & 0.681 & 0.579 & -0.186 & 0.987  & 0.163 & 0.093 & 0.765 & 0.287 \\
AR: 0.3 / CR: 0    & Equal  & 6 & 3 & 0.769 & 0.675 & -0.072 & 0.221  & 0.122 & 0.277 & 0.796 & 0.577 \\
AR: 0.3 / CR: 0    & Equal  & 6 & 7 & 0.8   & 0.755 & 0.06   & 0.884  & 0.122 & 0.192 & 0.778 & 0.472 \\
AR: 0.3 / CR: 0    & Offset & 3 & 3 & 0.625 & 0.474 & -0.158 & 0.939  & 0.183 & 0.114 & 0.765 & 0.4   \\
AR: 0.3 / CR: 0    & Offset & 3 & 7 & 0.68  & 0.572 & -0.173 & 0.971  & 0.167 & 0.091 & 0.807 & 0.29  \\
AR: 0.3 / CR: 0    & Offset & 6 & 3 & 0.695 & 0.411 & -0.128 & -0.152 & 0.152 & 0.253 & 0.808 & 0.412 \\
AR: 0.3 / CR: 0    & Offset & 6 & 7 & 0.789 & 0.669 & 0.005  & 0.61   & 0.114 & 0.239 & 0.746 & 0.574 \\
AR: 0.3 / CR: 0.25 & Equal  & 3 & 3 & 0.657 & 0.564 & -0.091 & 0.806  & 0.174 & 0.111 & 0.788 & 0.441 \\
AR: 0.3 / CR: 0.25 & Equal  & 3 & 7 & 0.69  & 0.61  & -0.078 & 0.801  & 0.165 & 0.095 & 0.816 & 0.373 \\
AR: 0.3 / CR: 0.25 & Equal  & 6 & 3 & 0.776 & 0.706 & -0.102 & 0.234  & 0.118 & 0.26  & 0.794 & 0.577 \\
AR: 0.3 / CR: 0.25 & Equal  & 6 & 7 & 0.803 & 0.764 & 0.047  & 0.735  & 0.122 & 0.183 & 0.772 & 0.522 \\
AR: 0.3 / CR: 0.25 & Offset & 3 & 3 & 0.625 & 0.491 & -0.247 & 0.765  & 0.181 & 0.125 & 0.766 & 0.49  \\
AR: 0.3 / CR: 0.25 & Offset & 3 & 7 & 0.69  & 0.594 & -0.098 & 0.827  & 0.171 & 0.098 & 0.835 & 0.397 \\
AR: 0.3 / CR: 0.25 & Offset & 6 & 3 & 0.698 & 0.419 & -0.137 & -0.238 & 0.151 & 0.256 & 0.769 & 0.399 \\
AR: 0.3 / CR: 0.25 & Offset & 6 & 7 & 0.797 & 0.688 & 0      & 0.519  & 0.115 & 0.242 & 0.749 & 0.59  \\
AR: 0.7 / CR: 0    & Equal  & 3 & 3 & 0.691 & 0.63  & -0.143 & 0.043  & 0.131 & 0.076 & 0.767 & 0.632 \\
AR: 0.7 / CR: 0    & Equal  & 3 & 7 & 0.731 & 0.666 & -0.083 & 0.074  & 0.122 & 0.062 & 0.775 & 0.535 \\
AR: 0.7 / CR: 0    & Equal  & 6 & 3 & 0.796 & 0.696 & -0.064 & -0.34  & 0.094 & 0.257 & 0.802 & 0.615 \\
AR: 0.7 / CR: 0    & Equal  & 6 & 7 & 0.826 & 0.785 & -0.009 & 0      & 0.083 & 0.161 & 0.786 & 0.59  \\
AR: 0.7 / CR: 0    & Offset & 3 & 3 & 0.671 & 0.568 & -0.139 & -0.007 & 0.14  & 0.094 & 0.782 & 0.648 \\
AR: 0.7 / CR: 0    & Offset & 3 & 7 & 0.724 & 0.664 & -0.119 & 0.079  & 0.127 & 0.063 & 0.794 & 0.535 \\
AR: 0.7 / CR: 0    & Offset & 6 & 3 & 0.737 & 0.443 & -0.099 & -0.671 & 0.114 & 0.226 & 0.766 & 0.332 \\
AR: 0.7 / CR: 0    & Offset & 6 & 7 & 0.808 & 0.714 & -0.024 & -0.2   & 0.083 & 0.226 & 0.758 & 0.674 \\
AR: 0.7 / CR: 0.25 & Equal  & 3 & 3 & 0.718 & 0.666 & -0.122 & -0.113 & 0.14  & 0.113 & 0.785 & 0.712 \\
AR: 0.7 / CR: 0.25 & Equal  & 3 & 7 & 0.752 & 0.706 & -0.091 & -0.079 & 0.139 & 0.086 & 0.8   & 0.627 \\
AR: 0.7 / CR: 0.25 & Equal  & 6 & 3 & 0.81  & 0.732 & -0.05  & -0.413 & 0.094 & 0.303 & 0.768 & 0.621 \\
AR: 0.7 / CR: 0.25 & Equal  & 6 & 7 & 0.833 & 0.788 & -0.021 & -0.168 & 0.092 & 0.213 & 0.773 & 0.721 \\
AR: 0.7 / CR: 0.25 & Offset & 3 & 3 & 0.693 & 0.616 & -0.129 & -0.148 & 0.154 & 0.116 & 0.787 & 0.681 \\
AR: 0.7 / CR: 0.25 & Offset & 3 & 7 & 0.751 & 0.699 & -0.09  & -0.067 & 0.134 & 0.094 & 0.828 & 0.642 \\
AR: 0.7 / CR: 0.25 & Offset & 6 & 3 & 0.755 & 0.497 & -0.105 & -0.644 & 0.121 & 0.248 & 0.762 & 0.358 \\
AR: 0.7 / CR: 0.25 & Offset & 6 & 7 & 0.828 & 0.739 & -0.034 & -0.314 & 0.09  & 0.247 & 0.78  & 0.715 \\ \hline
\end{tabular}%
}
\end{table}

\begin{table}[H]
\centering
\caption{State Recovery and Autoregressive Parameter Outcomes for 500 Timepoint Conditions. N Items: Number of Items per State; N Responses: Number of Response Categories per Item.}
\label{tab:ar500tps}
\resizebox{\columnwidth}{!}{%
\begin{tabular}{llcccccccccc}
\hline
AR/CR Parameters &
  Spread &
  N Items &
  \multicolumn{1}{c|}{N Response} &
  \multicolumn{2}{c|}{State Recovery} &
  \multicolumn{2}{c|}{Median Rel. Bias} &
  \multicolumn{2}{c|}{Median Slice SE} &
  \multicolumn{2}{c}{Coverage} \\
 &
   &
   &
  \multicolumn{1}{c|}{} &
  GRM &
  \multicolumn{1}{c|}{Linear} &
  GRM &
  \multicolumn{1}{c|}{Linear} &
  GRM &
  \multicolumn{1}{c|}{Linear} &
  GRM &
  Linear \\ \hline
AR: 0.3 / CR: 0    & Equal  & 3 & 3 & 0.658 & 0.482 & -0.073 & 1.16   & 0.067 & 0.046 & 0.77  & 0.091 \\
AR: 0.3 / CR: 0    & Equal  & 3 & 7 & 0.698 & 0.573 & -0.046 & 1.183  & 0.063 & 0.038 & 0.747 & 0.04  \\
AR: 0.3 / CR: 0    & Equal  & 6 & 3 & 0.777 & 0.593 & 0      & -0.139 & 0.047 & 0.089 & 0.714 & 0.207 \\
AR: 0.3 / CR: 0    & Equal  & 6 & 7 & 0.807 & 0.738 & 0.015  & 0.469  & 0.046 & 0.085 & 0.677 & 0.323 \\
AR: 0.3 / CR: 0    & Offset & 3 & 3 & 0.635 & 0.417 & -0.028 & 1.109  & 0.073 & 0.047 & 0.784 & 0.088 \\
AR: 0.3 / CR: 0    & Offset & 3 & 7 & 0.697 & 0.558 & -0.024 & 1.199  & 0.062 & 0.038 & 0.77  & 0.038 \\
AR: 0.3 / CR: 0    & Offset & 6 & 3 & 0.707 & 0.354 & -0.03  & -0.168 & 0.059 & 0.078 & 0.718 & 0     \\
AR: 0.3 / CR: 0    & Offset & 6 & 7 & 0.798 & 0.637 & 0.051  & 0.248  & 0.044 & 0.099 & 0.63  & 0.357 \\
AR: 0.3 / CR: 0.25 & Equal  & 3 & 3 & 0.663 & 0.527 & -0.054 & 0.971  & 0.068 & 0.049 & 0.778 & 0.135 \\
AR: 0.3 / CR: 0.25 & Equal  & 3 & 7 & 0.7   & 0.603 & -0.033 & 0.912  & 0.062 & 0.041 & 0.763 & 0.128 \\
AR: 0.3 / CR: 0.25 & Equal  & 6 & 3 & 0.78  & 0.631 & 0.01   & -0.022 & 0.047 & 0.094 & 0.71  & 0.244 \\
AR: 0.3 / CR: 0.25 & Equal  & 6 & 7 & 0.81  & 0.744 & 0.045  & 0.311  & 0.046 & 0.085 & 0.654 & 0.324 \\
AR: 0.3 / CR: 0.25 & Offset & 3 & 3 & 0.642 & 0.453 & -0.025 & 0.904  & 0.073 & 0.05  & 0.747 & 0.174 \\
AR: 0.3 / CR: 0.25 & Offset & 3 & 7 & 0.704 & 0.602 & -0.002 & 0.959  & 0.062 & 0.043 & 0.766 & 0.129 \\
AR: 0.3 / CR: 0.25 & Offset & 6 & 3 & 0.71  & 0.37  & -0.05  & -0.302 & 0.057 & 0.077 & 0.733 & 0     \\
AR: 0.3 / CR: 0.25 & Offset & 6 & 7 & 0.801 & 0.655 & 0.028  & 0.042  & 0.045 & 0.095 & 0.68  & 0.309 \\
AR: 0.7 / CR: 0    & Equal  & 3 & 3 & 0.717 & 0.661 & -0.028 & 0.182  & 0.046 & 0.024 & 0.728 & 0.168 \\
AR: 0.7 / CR: 0    & Equal  & 3 & 7 & 0.748 & 0.702 & -0.012 & 0.188  & 0.042 & 0.019 & 0.736 & 0.109 \\
AR: 0.7 / CR: 0    & Equal  & 6 & 3 & 0.81  & 0.635 & -0.019 & -0.475 & 0.034 & 0.089 & 0.718 & 0     \\
AR: 0.7 / CR: 0    & Equal  & 6 & 7 & 0.836 & 0.767 & 0.016  & -0.148 & 0.03  & 0.087 & 0.63  & 0.255 \\
AR: 0.7 / CR: 0    & Offset & 3 & 3 & 0.697 & 0.582 & -0.033 & 0.147  & 0.049 & 0.027 & 0.767 & 0.262 \\
AR: 0.7 / CR: 0    & Offset & 3 & 7 & 0.748 & 0.701 & -0.019 & 0.191  & 0.043 & 0.019 & 0.732 & 0.112 \\
AR: 0.7 / CR: 0    & Offset & 6 & 3 & 0.755 & 0.398 & -0.02  & -0.694 & 0.041 & 0.081 & 0.761 & 0.097 \\
AR: 0.7 / CR: 0    & Offset & 6 & 7 & 0.826 & 0.677 & 0.012  & -0.366 & 0.031 & 0.099 & 0.646 & 0.294 \\
AR: 0.7 / CR: 0.25 & Equal  & 3 & 3 & 0.743 & 0.689 & -0.038 & -0.004 & 0.05  & 0.038 & 0.737 & 0.5   \\
AR: 0.7 / CR: 0.25 & Equal  & 3 & 7 & 0.773 & 0.738 & -0.041 & 0.034  & 0.047 & 0.031 & 0.719 & 0.48  \\
AR: 0.7 / CR: 0.25 & Equal  & 6 & 3 & 0.823 & 0.681 & -0.012 & -0.554 & 0.035 & 0.097 & 0.72  & 0.158 \\
AR: 0.7 / CR: 0.25 & Equal  & 6 & 7 & 0.849 & 0.79  & 0.003  & -0.312 & 0.032 & 0.092 & 0.671 & 0.287 \\
AR: 0.7 / CR: 0.25 & Offset & 3 & 3 & 0.73  & 0.63  & -0.039 & -0.033 & 0.053 & 0.039 & 0.731 & 0.4   \\
AR: 0.7 / CR: 0.25 & Offset & 3 & 7 & 0.773 & 0.735 & -0.031 & 0.013  & 0.046 & 0.032 & 0.756 & 0.492 \\
AR: 0.7 / CR: 0.25 & Offset & 6 & 3 & 0.775 & 0.423 & -0.027 & -0.708 & 0.044 & 0.082 & 0.721 & 0     \\
AR: 0.7 / CR: 0.25 & Offset & 6 & 7 & 0.84  & 0.73  & 0.002  & -0.479 & 0.032 & 0.104 & 0.704 & 0     \\ \hline
\end{tabular}%
}
\end{table}

\begin{table}[H]
\centering
\caption{Cross Regressive Parameter Outcomes for 100 Timepoint Conditions. N Items: Number of Items per State; N Responses: Number of Response Categories per Item.}
\label{tab:cr100tps}
\resizebox{\columnwidth}{!}{%
\begin{tabular}{llcccccccc}
\hline
AR/CR Parameters &
  Spread &
  N Items &
  \multicolumn{1}{c|}{N Responses} &
  \multicolumn{2}{c|}{Median Bias} &
  \multicolumn{2}{c|}{Median Slice SE} &
  \multicolumn{2}{c}{Coverage} \\
 &
   &
   &
  \multicolumn{1}{c|}{} &
  GRM &
  \multicolumn{1}{c|}{Linear} &
  GRM &
  \multicolumn{1}{c|}{Linear} &
  GRM &
  Linear \\ \hline
AR: 0.3 / CR: 0    & Equal  & 3 & 3 & -0.005 & 0.42  & 0.16  & 0.11  & 0.823 & 0.204 \\
AR: 0.3 / CR: 0    & Equal  & 3 & 7 & 0.008  & 0.371 & 0.164 & 0.093 & 0.82  & 0.143 \\
AR: 0.3 / CR: 0    & Equal  & 6 & 3 & 0.008  & 0.489 & 0.116 & 0.205 & 0.778 & 0.411 \\
AR: 0.3 / CR: 0    & Equal  & 6 & 7 & 0.003  & 0.395 & 0.114 & 0.155 & 0.803 & 0.36  \\
AR: 0.3 / CR: 0    & Offset & 3 & 3 & -0.024 & 0.428 & 0.18  & 0.115 & 0.8   & 0.193 \\
AR: 0.3 / CR: 0    & Offset & 3 & 7 & -0.02  & 0.381 & 0.162 & 0.092 & 0.85  & 0.147 \\
AR: 0.3 / CR: 0    & Offset & 6 & 3 & 0.015  & 0.367 & 0.147 & 0.204 & 0.814 & 0.337 \\
AR: 0.3 / CR: 0    & Offset & 6 & 7 & 0.008  & 0.424 & 0.109 & 0.172 & 0.759 & 0.388 \\
AR: 0.3 / CR: 0.25 & Equal  & 3 & 3 & 0.047  & 0.316 & 0.164 & 0.119 & 0.753 & 0.365 \\
AR: 0.3 / CR: 0.25 & Equal  & 3 & 7 & 0.035  & 0.323 & 0.158 & 0.108 & 0.823 & 0.324 \\
AR: 0.3 / CR: 0.25 & Equal  & 6 & 3 & 0.016  & 0.279 & 0.113 & 0.213 & 0.783 & 0.548 \\
AR: 0.3 / CR: 0.25 & Equal  & 6 & 7 & -0.011 & 0.244 & 0.112 & 0.153 & 0.774 & 0.591 \\
AR: 0.3 / CR: 0.25 & Offset & 3 & 3 & 0.029  & 0.264 & 0.183 & 0.124 & 0.763 & 0.431 \\
AR: 0.3 / CR: 0.25 & Offset & 3 & 7 & 0.014  & 0.305 & 0.165 & 0.102 & 0.827 & 0.367 \\
AR: 0.3 / CR: 0.25 & Offset & 6 & 3 & -0.011 & 0.185 & 0.146 & 0.227 & 0.777 & 0.385 \\
AR: 0.3 / CR: 0.25 & Offset & 6 & 7 & 0.014  & 0.274 & 0.11  & 0.198 & 0.734 & 0.63  \\
AR: 0.7 / CR: 0    & Equal  & 3 & 3 & 0      & 0.273 & 0.11  & 0.078 & 0.743 & 0.157 \\
AR: 0.7 / CR: 0    & Equal  & 3 & 7 & 0      & 0.248 & 0.098 & 0.061 & 0.783 & 0.133 \\
AR: 0.7 / CR: 0    & Equal  & 6 & 3 & 0.006  & 0.342 & 0.081 & 0.197 & 0.736 & 0.492 \\
AR: 0.7 / CR: 0    & Equal  & 6 & 7 & -0.017 & 0.225 & 0.077 & 0.117 & 0.779 & 0.559 \\
AR: 0.7 / CR: 0    & Offset & 3 & 3 & -0.007 & 0.289 & 0.121 & 0.084 & 0.753 & 0.213 \\
AR: 0.7 / CR: 0    & Offset & 3 & 7 & -0.013 & 0.252 & 0.112 & 0.062 & 0.829 & 0.151 \\
AR: 0.7 / CR: 0    & Offset & 6 & 3 & 0.006  & 0.318 & 0.094 & 0.211 & 0.736 & 0.354 \\
AR: 0.7 / CR: 0    & Offset & 6 & 7 & 0.002  & 0.353 & 0.071 & 0.164 & 0.698 & 0.565 \\
AR: 0.7 / CR: 0.25 & Equal  & 3 & 3 & 0.053  & 0.202 & 0.115 & 0.118 & 0.783 & 0.533 \\
AR: 0.7 / CR: 0.25 & Equal  & 3 & 7 & 0.067  & 0.204 & 0.113 & 0.095 & 0.823 & 0.43  \\
AR: 0.7 / CR: 0.25 & Equal  & 6 & 3 & 0.055  & 0.227 & 0.077 & 0.23  & 0.695 & 0.617 \\
AR: 0.7 / CR: 0.25 & Equal  & 6 & 7 & 0.027  & 0.164 & 0.075 & 0.165 & 0.716 & 0.732 \\
AR: 0.7 / CR: 0.25 & Offset & 3 & 3 & 0.074  & 0.231 & 0.127 & 0.124 & 0.795 & 0.456 \\
AR: 0.7 / CR: 0.25 & Offset & 3 & 7 & 0.075  & 0.246 & 0.106 & 0.106 & 0.776 & 0.425 \\
AR: 0.7 / CR: 0.25 & Offset & 6 & 3 & 0.062  & 0.281 & 0.095 & 0.218 & 0.705 & 0.337 \\
AR: 0.7 / CR: 0.25 & Offset & 6 & 7 & 0.018  & 0.216 & 0.072 & 0.201 & 0.712 & 0.667 \\ \hline
\end{tabular}%
}
\end{table}

\begin{table}[H]
\centering
\caption{Cross Regressive Parameter Outcomes for 500 Timepoint Conditions. N Items: Number of Items per State; N Responses: Number of Response Categories per Item.}
\label{tab:cr500tps}
\resizebox{\columnwidth}{!}{%
\begin{tabular}{llcccccccc}
\hline
AR/CR Parameters &
  Spread &
  N Items &
  \multicolumn{1}{c|}{N Responses} &
  \multicolumn{2}{c|}{Median Bias} &
  \multicolumn{2}{c|}{Median Slice SE} &
  \multicolumn{2}{c}{Coverage} \\
 &
   &
   &
  \multicolumn{1}{c|}{} &
  GRM &
  \multicolumn{1}{c|}{Linear} &
  GRM &
  \multicolumn{1}{c|}{Linear} &
  GRM &
  Linear \\ \hline
AR: 0.3 / CR: 0    & Equal  & 3 & 3 & -0.006 & 0.331  & 0.065 & 0.043 & 0.715 & 0.037 \\
AR: 0.3 / CR: 0    & Equal  & 3 & 7 & -0.002 & 0.365  & 0.06  & 0.036 & 0.721 & 0.01  \\
AR: 0.3 / CR: 0    & Equal  & 6 & 3 & 0.004  & 0.476  & 0.048 & 0.072 & 0.764 & 0.111 \\
AR: 0.3 / CR: 0    & Equal  & 6 & 7 & -0.001 & 0.434  & 0.045 & 0.059 & 0.708 & 0.05  \\
AR: 0.3 / CR: 0    & Offset & 3 & 3 & -0.005 & 0.312  & 0.069 & 0.045 & 0.762 & 0.051 \\
AR: 0.3 / CR: 0    & Offset & 3 & 7 & 0.001  & 0.353  & 0.06  & 0.036 & 0.74  & 0.01  \\
AR: 0.3 / CR: 0    & Offset & 6 & 3 & 0.008  & 0.292  & 0.055 & 0.079 & 0.758 & 0     \\
AR: 0.3 / CR: 0    & Offset & 6 & 7 & 0.003  & 0.465  & 0.041 & 0.062 & 0.693 & 0.073 \\
AR: 0.3 / CR: 0.25 & Equal  & 3 & 3 & -0.012 & 0.167  & 0.064 & 0.05  & 0.727 & 0.168 \\
AR: 0.3 / CR: 0.25 & Equal  & 3 & 7 & 0.005  & 0.266  & 0.059 & 0.041 & 0.765 & 0.124 \\
AR: 0.3 / CR: 0.25 & Equal  & 6 & 3 & -0.004 & 0.054  & 0.046 & 0.075 & 0.72  & 0.198 \\
AR: 0.3 / CR: 0.25 & Equal  & 6 & 7 & -0.016 & 0.284  & 0.043 & 0.06  & 0.697 & 0.157 \\
AR: 0.3 / CR: 0.25 & Offset & 3 & 3 & 0.005  & 0.228  & 0.07  & 0.051 & 0.811 & 0.179 \\
AR: 0.3 / CR: 0.25 & Offset & 3 & 7 & 0.006  & 0.285  & 0.058 & 0.046 & 0.726 & 0.144 \\
AR: 0.3 / CR: 0.25 & Offset & 6 & 3 & 0.014  & -0.109 & 0.057 & 0.065 & 0.748 & 0     \\
AR: 0.3 / CR: 0.25 & Offset & 6 & 7 & -0.009 & 0.274  & 0.042 & 0.068 & 0.629 & 0.169 \\
AR: 0.7 / CR: 0    & Equal  & 3 & 3 & -0.003 & 0.165  & 0.036 & 0.02  & 0.742 & 0.03  \\
AR: 0.7 / CR: 0    & Equal  & 3 & 7 & 0.002  & 0.181  & 0.034 & 0.018 & 0.796 & 0.007 \\
AR: 0.7 / CR: 0    & Equal  & 6 & 3 & 0      & 0.256  & 0.03  & 0.069 & 0.711 & 0     \\
AR: 0.7 / CR: 0    & Equal  & 6 & 7 & -0.003 & 0.278  & 0.028 & 0.055 & 0.69  & 0.066 \\
AR: 0.7 / CR: 0    & Offset & 3 & 3 & -0.003 & 0.16   & 0.037 & 0.024 & 0.765 & 0.075 \\
AR: 0.7 / CR: 0    & Offset & 3 & 7 & -0.004 & 0.165  & 0.035 & 0.016 & 0.773 & 0.003 \\
AR: 0.7 / CR: 0    & Offset & 6 & 3 & 0.003  & 0.205  & 0.031 & 0.066 & 0.71  & 0.125 \\
AR: 0.7 / CR: 0    & Offset & 6 & 7 & 0.004  & 0.319  & 0.027 & 0.065 & 0.709 & 0.126 \\
AR: 0.7 / CR: 0.25 & Equal  & 3 & 3 & 0.011  & 0.096  & 0.038 & 0.041 & 0.747 & 0.191 \\
AR: 0.7 / CR: 0.25 & Equal  & 3 & 7 & 0.018  & 0.13   & 0.036 & 0.033 & 0.714 & 0.182 \\
AR: 0.7 / CR: 0.25 & Equal  & 6 & 3 & -0.003 & -0.01  & 0.028 & 0.076 & 0.729 & 0.188 \\
AR: 0.7 / CR: 0.25 & Equal  & 6 & 7 & 0.003  & 0.193  & 0.025 & 0.063 & 0.615 & 0.288 \\
AR: 0.7 / CR: 0.25 & Offset & 3 & 3 & 0.022  & 0.114  & 0.041 & 0.04  & 0.731 & 0.115 \\
AR: 0.7 / CR: 0.25 & Offset & 3 & 7 & 0.012  & 0.167  & 0.035 & 0.037 & 0.67  & 0.175 \\
AR: 0.7 / CR: 0.25 & Offset & 6 & 3 & 0.013  & -0.054 & 0.032 & 0.077 & 0.711 & 0.084 \\
AR: 0.7 / CR: 0.25 & Offset & 6 & 7 & 0.001  & 0.232  & 0.025 & 0.071 & 0.577 & 0.259 \\ \hline
\end{tabular}%
}
\end{table}

\section{Slice Likelihood Standard Error Figures}
\renewcommand{\thefigure}{S\arabic{figure}}
\setcounter{figure}{0}

\begin{figure}[H]
    \centering
    \includegraphics[width=.99\textwidth]{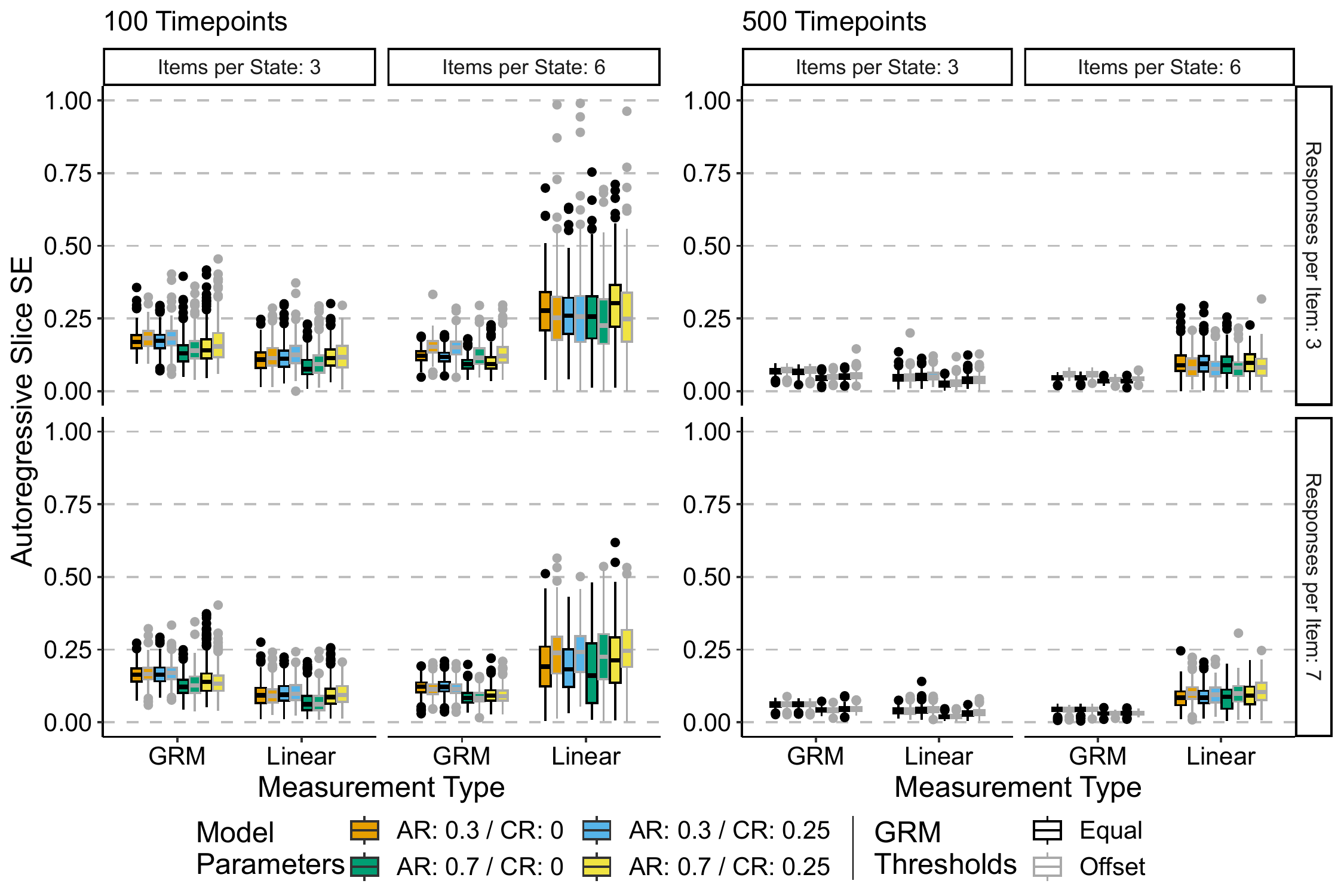}
    \caption{Autoregressive Parameter Slice SE. Central black line on boxplot denotes median, box denotes $25\%-75\%$ interquartile range, whiskers denote $1.5 \times IQR$. AR refers to the autoregressive parameter while CR refers to the cross-regressive parameter.}
    \label{fig:arse}
\end{figure}

\begin{figure}[H]
    \centering
    \includegraphics[width=.99\textwidth]{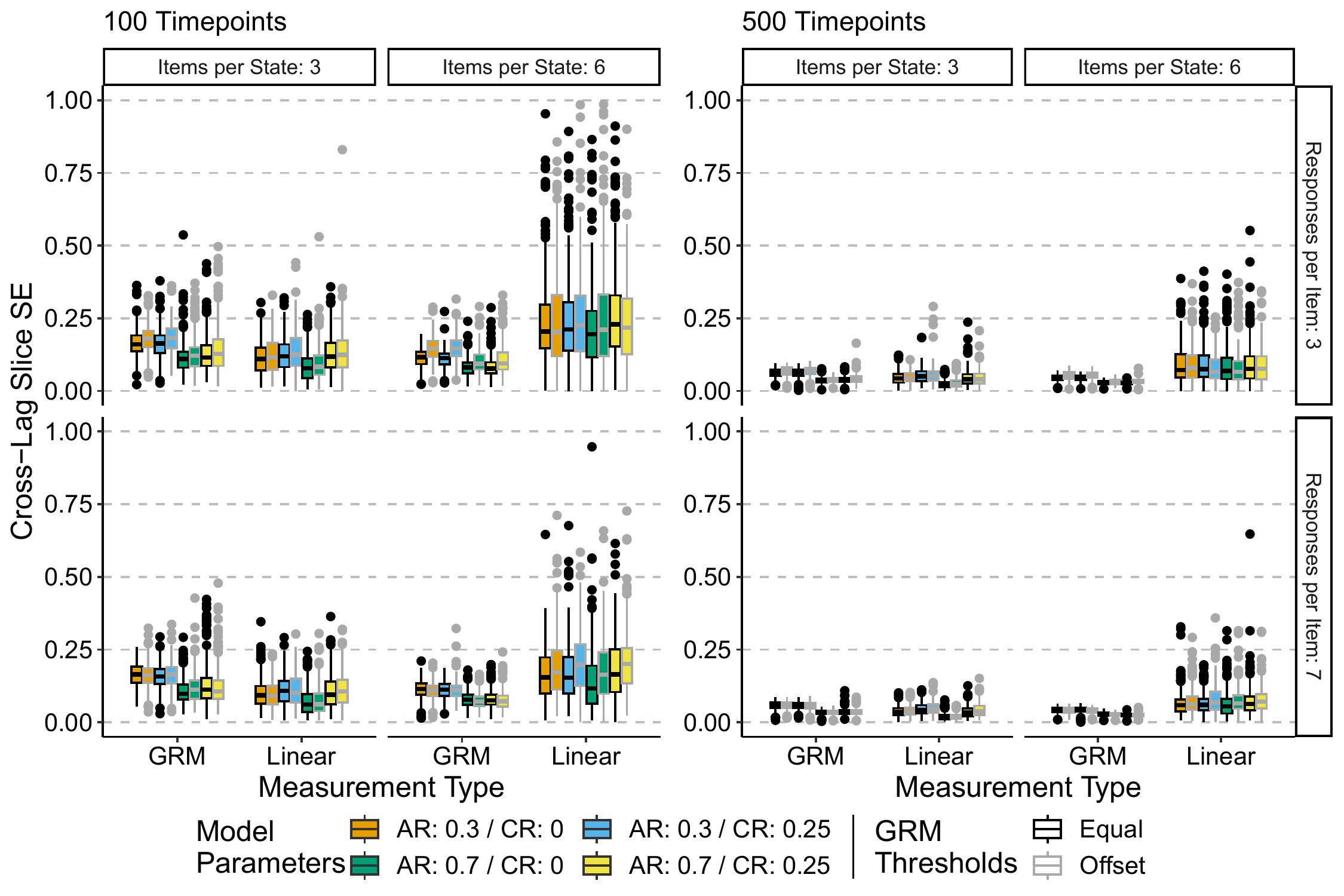}
    \caption{Cross-regressive Slice SE. Central black line on boxplot denotes median, box denotes $25\%-75\%$ interquartile range, whiskers denote $1.5 \times IQR$. AR refers to the autoregressive parameter while CR refers to the cross-regressive parameter.}
    \label{fig:crse}
\end{figure}